\documentclass{article}

\usepackage{arxiv}

\usepackage[utf8]{inputenc} 
\usepackage[T1]{fontenc}    
\usepackage{hyperref}       
\usepackage{url}            
\usepackage{booktabs}       
\usepackage{amsfonts}       
\usepackage{nicefrac}       
\usepackage{microtype}      
\usepackage{lipsum}
\usepackage{graphicx}
\graphicspath{ {./images/} }

\usepackage{subcaption}
\usepackage{threeparttable}
\usepackage{commath}
\usepackage{amsthm}
\usepackage{amsfonts}
\usepackage{multirow}
\usepackage{booktabs}
\usepackage{xcolor}
\usepackage{subcaption}
\usepackage{float}
\usepackage{amsmath}
\usepackage{amssymb}

\theoremstyle{definition}
\newtheorem{definition}{Definition}[section]
\newtheorem{theorem}{Theorem}[section]
\newtheorem{lemma}[theorem]{Lemma}
\usepackage[linesnumbered,ruled]{algorithm2e}

\title{DP-Image: Differential Privacy for Image Data in Feature Space}

\author{
 Hanyu Xue \\
  School of Computer Science\\
  University of Technology Sydney\\
  Australia \\
  \texttt{hanyu.xue@student.uts.edu.au} \\
    \And
 Bo Liu \\
  School of Computer Science\\
  University of Technology Sydney\\
  Australia \\
  \texttt{bo.liu@uts.edu.au} \\
   \And
 Ming Ding \\
  Data61\\
  CSIRO \\
  Australia \\
  \texttt{ming.ding@data61.csiro.au} \\
  \And
 Tianqing Zhu \\
  School of Computer Science\\
  University of Technology Sydney\\
  Australia \\
  \texttt{tianqing.zhu@uts.edu.au} \\
    \And
  Dayong Ye \\
  School of Computer Science\\
  University of Technology Sydney\\
  Australia \\
  \texttt{Dayong.Ye@uts.edu.au} \\
  \And
    Li Song \\
  Shanghai Jiao Tong University\\
  Department of Electronic Engineering\\
  China \\
  \texttt{song\_li@sjtu.edu.cn} \\
     \And
  Wanlei Zhou \\
  City University of Macau\\
  \texttt{wlzhou@cityu.mo} \\
}

\begin{document}
\maketitle
\begin{abstract}
{The excessive use of images in social networks, government databases, and industrial applications has posed great privacy risks and raised serious concerns from the public. Even though differential privacy (DP) is a widely accepted criterion that can provide a provable privacy guarantee, the application of DP on unstructured data such as images is not trivial due to the lack of a clear qualification on the meaningful difference between any two images. In this paper, for the first time, we introduce a novel notion of image aware differential privacy, referred to as DP-image, that can protect user's personal information in images, from both human and AI adversaries. The DP-Image definition is formulated as an extended version of traditional differential privacy, considering the distance measurements between feature space vectors of images. Then we propose a mechanism to achieve DP-Image by adding noise to an image feature vector. Finally, we conduct experiments with a case study on face image privacy. Our results show that the proposed DP-Image method provides excellent DP protection on images, with a controllable distortion to faces.}
\end{abstract}


\section{Introduction}
\label{sec:Intro}
Recently, 
our human society witnesses a rapid increase in the use of image data,
which poses high risks of privacy leakage, 
as images usually contain a large number of sensitive data that might visually reveal personal information~\cite{facebookprivacy}~\cite{liu2021ACM}. 
For example, 
heavy concerns on the privacy risks were raised on uploading people's face pictures to a popular APP called FaceApp~\cite{Faceapp}, 
which can edit a person's face by changing his/her gender, age, ethnicity, etc.
In fact, 
the instinct of protecting privacy in image data is not new in our human society.
So the question is: 
\emph{why the issue of image privacy becomes urgent now?}
This is because the newly emerging deep learning techniques have exacerbated the privacy risks for image sharing and usage. 
In more detail,
artificial intelligence (AI) aided by deep learning methods can automatically detect and collect people's private and sensitive information, 
thus impacting on everyone's daily life.
This brings risks of personal privacy to a whole new level, 
while the traditional privacy-preserving methods seem powerless when facing the large-scale deep learning tools~\cite{mcpherson2016defeating}. 

In this light, 
privacy issues have become hot topics in government debates and legislation, 
as many countries have launched privacy acts and laws. 
For example, 
the European General Data Protection Regulation (GDPR)~\cite{GDPR} took effect on 25 May 2018. 
It emphasizes the protection of ``personal data'', 
interpreted as ``any information relating to an identified or identifiable natural person (‘data subject’); 
an identifiable natural person is one who can be identified, 
directly or indirectly, 
in particular by reference to an identifier such as a name, 
an identification number, 
location data, 
an online identifier or to one or more factors specific to the physical,
physiological, genetic, mental, economic, 
cultural or social identity of that natural person''. 
According to this definition, 
images contain a variety of personal identifiers such as people's faces, text, and license plates. 
Therefore, 
effective privacy protection methods for the image data are in urgent need.

Unfortunately, 
privacy protection for unstructured data, 
such as images, 
is much more difficult compared with that for structured data. 
Structured data are usually well documented in an organized and systematic manner. 
However, 
in unstructured data such as images, 
data attributes are implicitly represented by sets of pixels covering irregular shapes and sizes. 
Such implicit data attribute values cannot be processed by traditional statistical methods, 
such as correlation, regression, etc.  
Hence, 
the underlying privacy risks in unstructured data are much less clear than those in a structured one.

Due to the unstructured nature of the image data, 
research on image privacy protection takes a different approach compared with that on structured data. 
At first, 
traditional research on image privacy protection often assumes \emph{human adversaries}.
In other words, 
the privacy risks are usually quantified by how effectively the information contained in images can be picked up by human eyes and brains.
As a result, 
``blurring'', ``pixelation'', and ``mosaic'' are the most used methods to protect privacy in images, 
e.g. in Google Earth street view. 
However, 
we are now marching into a new age of AI, 
where the information contained in images can be filtered out by AI functions/models. 
Those AI functions such as DNNs seem to take a different approach to interpret and understand images, 
and they are able to re-identify the obfuscated image with high accuracy~\cite{mcpherson2016defeating}. 
Therefore, 
very recently, 
researchers have started to consider a new scenario, 
where AI acts as an adversary. 
For example, 
to throw off AI adversaries when analyzing images,
the authors of~\cite{goodfellow2014explaining} proposed to generate a small but intentional worst-case disturbance to an original image, which can mislead DNNs in machine learning tasks such as image classification, 
without causing a significant visual difference perceptible to human eyes. 
The perturbed image is referred to as an ``adversarial example'', 
and the specially generated noise pattern is thus named adversarial perturbations (AP). 
A few papers have discussed the potential of AP in privacy protection~\cite{liu2019adversaries, oh2017adversarial,liu2017protecting,liu19bmsb, shan2020fawkes, li2019anonymousnet, rajabi2021practicality}. 
Another recent trend for image privacy protection is the use of the generative adversarial network (GAN). 
Content generated by GAN can be used to replace or edit the original images so that the identity-related information can be removed~\cite{sun2018natural, wu2019privacy, sun2018hybrid, li2019faceshifter, wang2020infoscrub}. Finally, there have been several preliminary papers that apply the differential privacy (DP) in images. Fan proposed $\epsilon$- differential private methods in the pixel level of the image~\cite{fan2018image}, and in Singular Value Decomposition (SVD)~\cite{fan2019practical}, respectively. However, making image pixels or SVD indistinguishable does not make much sense in practice, and the quality of the generated image is quite low. Lecuyer et al.~\cite{lecuyer2019certified} proposed the PixelDP framework that includes a DP noise layer in the DNN. PixelDP scheme enforces that the output prediction function is DP provided the
input changes on a small number of pixels (when the input is an image). But the purpose of PixelDP is to increase robustness to adversarial examples, other than image privacy.  

Overall, 
the aforementioned works cannot solve the image privacy protection problem thoroughly due to several reasons: 
\begin{itemize}
    \item First, the image privacy is yet neither clearly defined, nor can it be quantitatively measured.
    \item Second, the current AP-based method is only designed for specific models and applications. It is not effective to human adversaries due to the uncontrollable visual appearance.
    \item Third, the GAN-based image manipulation lacks provable privacy measurements.
    \item Finally, the existing DP methods do not effectively capture the key features of images from the privacy perspective, or was not designed for the purpose of image privacy protection.
\end{itemize}

Motivated by these challenges, 
we propose a differentially private image (DP-Image) framework and design DP-Image mechanisms in this work. 
This framework redefines the traditional DP~\cite{dwork2008differential} and R\'enyi differential privacy (RDP)~\cite{mironov2017renyi} notions, in the context of the image data. 
It should be noted that the application of DP on the image data is not a trivial extension, 
due to the vastly different application scenarios. 
In traditional database applications, 
we assume that a data curator has a database $D$ and an attacker can either make a query on the database to obtain useful information to launch privacy attacks (i.e., the interactive setup) 
or get access to a synthetic version of database $D$ (i.e., the non-interactive setup). 
The important thing is that the adversary does NOT have access to the original database. 
Therefore, 
applying DP to the query answers or the synthetic data release will be able to protect the original data. 

However, 
the most common application of images is \textit{image publication and sharing}. 
It includes the case of personal image sharing on social network platforms and commercial image applications such as Google Street View. In this scenario, 
the original image data need to be exposed to some extent for meaningful applications. 
For example, 
it would be weird to share a synthetic photo of a cartoon character in a person's Facebook post, 
saying he/she took a selfie on a sunny Saturday afternoon. 
Hence, 
in the case of the image data, 
the adversary gains a considerable advantage in stepping closer to the original data than the conventional scenarios, 
and thus it becomes less difficult for the adversary to re-identify and collect sensitive information from published images. 
Therefore, 
the perturbation of privacy protection needs to be added directly onto the original image, 
so as to mislead the adversary when he analyzes the image. 
To this end, 
we propose a DP-Image framework that adds DP noise to the image feature vector in the latent space to alter the meaningful information carried by the image, 
and then reconstructs the image from the perturbed vector to replace the original image.
In this way, 
the reconstructed one will not disrupt the intended applications because it shares non-private information with the original one, 
which keeps the application context meaningful and consistent. 

\begin{figure*}[ht]
\begin{subfigure}{.195\textwidth}
  \centering
  \includegraphics[width=.9\linewidth]{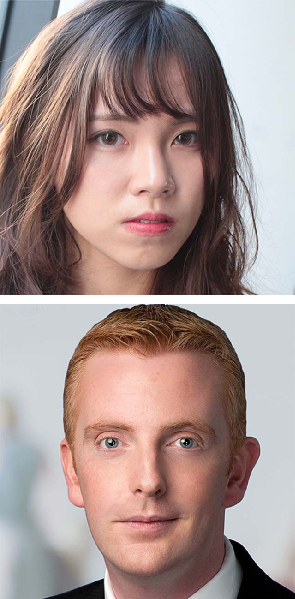} 
  \caption{\centering Original}
  \label{fig:intro1}
\end{subfigure}
\begin{subfigure}{.195\textwidth}
  \centering
  \includegraphics[width=.9\linewidth]{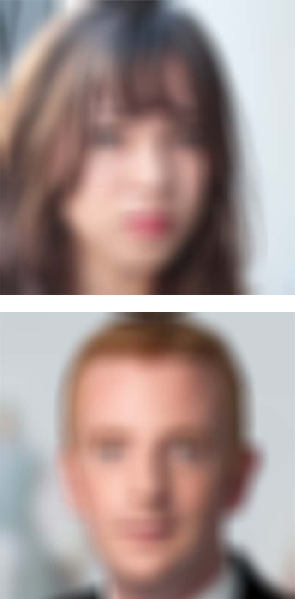}  
  \caption{\centering Blur}
  \label{fig:intro2}
\end{subfigure}
\begin{subfigure}{.195\textwidth}
  \centering
  \includegraphics[width=.9\linewidth]{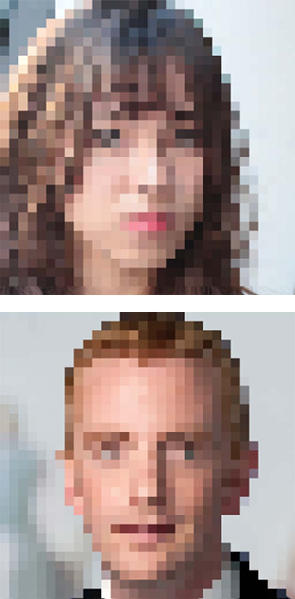}  
  \caption{\centering Mosaic}
  \label{fig:intro3}
\end{subfigure}
\begin{subfigure}{.195\textwidth}
  \centering
  \includegraphics[width=.9\linewidth]{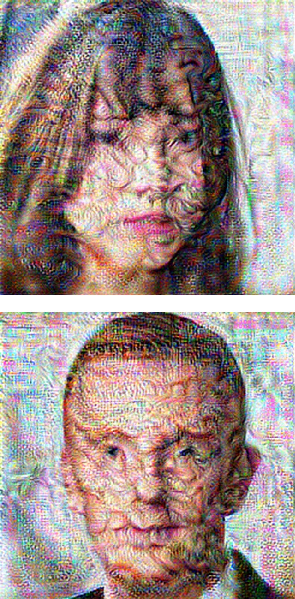}  
  \caption{\centering Adversarial(pixel)}
  \label{fig:intro4}
\end{subfigure}
\begin{subfigure}{.195\textwidth}
  \centering
  \includegraphics[width=.9\linewidth]{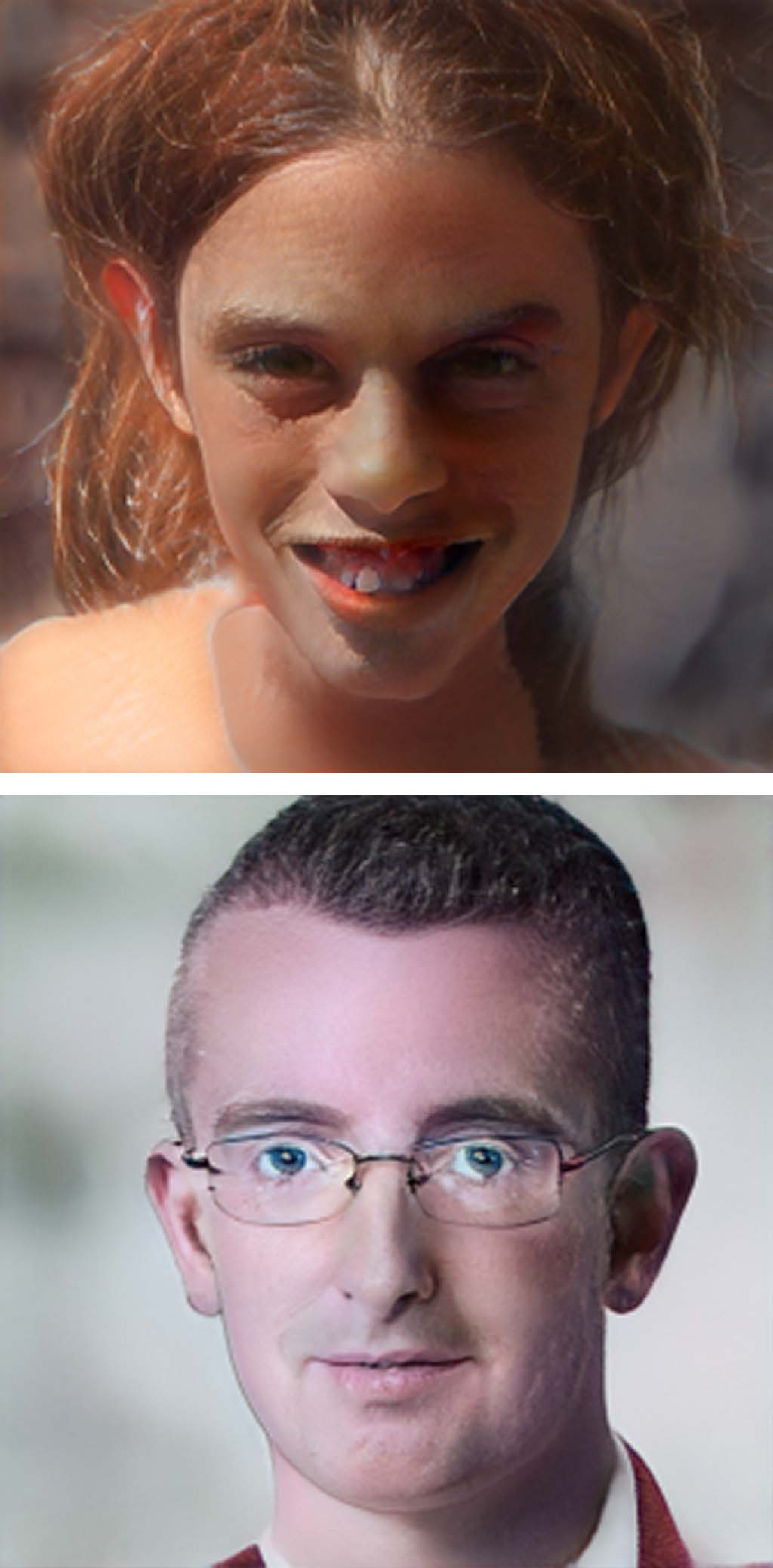}  
  \caption{\centering Our}
  \label{fig:intro5}
\end{subfigure}
\caption{A brief comparison of different image privacy protection methods. Our
approach not only protects the privacy of an image by generating a
photo-realistic alternative, but also provides a controllable way for privacy preservation.}
\label{fig: Introduction}
\end{figure*}

To elaborate on this in a more concrete manner, 
we show a comparison of our method with several traditional methods in Fig.~\ref{fig: Introduction}. As can be seen from the figure, 
the traditional methods such as blur and mosaic considerably destroy the utility of the face images. 
If such images were used in, e.g., demographics analysis, 
the application would be most likely to be unfruitful. 
Looking at the DP images generated by our method, 
we can see that the ones with unchanged identity are still useful for a possible demographics analysis or a straightforward Facebook post. 
Only some of the private information such as haircut, clothing, background location have been changed. And it can prevent human adversaries with a basic level of privacy protection.
Further, 
the images with changed identities provide even stronger privacy protection. 
Only the necessary information of gender and age for certain applications remains unchanged, 
the other private information is untraceable, 
which greatly protects an individual's privacy, against both human and AI adversaries. 

The major contributions of this paper are summarized as follows:
\begin{itemize}
    \item We propose a DP-Image definition based on the notions of DP and RDP. 
    \item We propose a DP-Image protection mechanism with provable and adjustable privacy levels.
    \item We design a DP-Image framework that enables us to apply the DP-Image protection mechanism in the image feature level, 
    utilizing advanced deep learning and GAN techniques.
    \item We implement the proposed DP-Image protection mechanisms on a real-life image dataset and show its effectiveness in safeguarding people's privacy.
\end{itemize}

The structure of the remainder of the paper is as follows. 
In Section \ref{sec:Preliminaries}, 
we discuss the preliminary knowledge for our work, 
including DP, RDP, and deep learning in image applications. 
In Section \ref{sec:System-Model}, 
we first define the adversary model. 
Then, 
we proceed to formulate the notion of DP-Image and present our DP-Image protection mechanisms.
In Section \ref{sec:experiments}, 
we conduct experiments to validate our proposed scheme, 
using a case study on face image privacy. We analyze the privacy guarantee of our proposed framework and schemes in Section \ref{sec:discussion}.
Section \ref{sec: related work} discusses the related work, 
and Section \ref{sec: conclusion} concludes our work.

\section{Preliminaries}
\label{sec:Preliminaries}

\subsection{Differential Privacy}
Informally, 
DP~\cite{dwork2008differential,Dwork2006calibrating} is a definition of privacy which guarantees that the likelihood of seeing an output on a given original dataset 
is close to the likelihood of seeing the same output on another dataset that differs from the original one in any single row. 
Here, an output could be another dataset, a statistical summary table, 
or a simple answer to a query, etc.
Since this single row is arbitrary, 
this definition aims to make an adversary unable to be certain about whether a particular individual is in the original dataset or not. 
In other words, 
differential privacy provides any individual in the dataset with plausible deniability
– the ability to say “I am not in that original dataset” -
and hence an individual can claim that he/she is not in the output. 
This is supported by mathematical proofs showing an employed method does provide the DP property.
Generally speaking, 
the basic idea of a DP mechanism is to introduce randomness into the original dataset, 
so that any individual’s information cannot be inferred by an adversary looking at the released output.

A key property of DP is that we can add outputs of several DP algorithms, 
and the result is still DP. 
However, 
this property incurs a loss in the privacy budget. 
The maximum impact is the sum of the privacy budgets for the involved outputs. 
By appropriately setting the privacy budget of constituent algorithms, 
we can ensure that the overall privacy budget of the data release algorithm is within the bound of a target privacy budget. 

Another property of DP is that if the dataset is divided into disjoint subsets,
such that addition or removal of a row affects at most one subset, 
then we can assign each dataset a privacy budget of $\epsilon$, 
and still maintain an overall privacy budget of $\epsilon$. 
These two properties are referred to as the sequential and parallel composition of DP.

A final useful property of DP is that the output of a DP algorithm can be post-processed without affecting privacy provided it is done independently of the original dataset. 
For example, 
averaging, rounding or any change to the numbers will not impact the privacy of the data. 
This means that an analyst can do any amount of data post-processing on a released DP dataset and cannot reduce its privacy guarantee. 

The DP treatment usually involves adding noise to give uncertainty to the impact of any single individual. 
We need to add noise in a scale relative to how much any individual could make a difference.
For example,
if we add noise to counts of individuals, 
then any single individual only influences that count by a maximum value of 1. 
So, 
the noise is on a scale relative to that 1. 

Following the above brief discussion on DP, here we show the formal definition of DP. 

\begin{definition}{\label{def:DP}\textbf{($\epsilon$-Differential Privacy) }~\cite{dwork2008differential,Dwork2006calibrating}:}
A randomized mechanism $\mathcal{M}$ gives $\epsilon$-differential privacy if for any neighboring datasets $D$ and $D^\prime$ differing on one element, and all sets of output $S$:
\begin{align}
Pr[\mathcal{M}(D)\in S] \leq \exp(\epsilon)\cdot Pr[\mathcal{M}(D^\prime)\in S].
\end{align}
\end{definition}
A random perturbation can be added to achieve differential privacy. 
Sensitivity calibrates the amount of noise for a specified query $f$ of dataset $D$. 
$\Delta f$ is the $l_1$-norm sensitivity defined as:
\begin{definition}{\label{def: sensitivity}\textbf{($l_1-$norm sensitivity) }\cite{dwork2008differential}:}
For any query $f$: $D \to \mathbb{R}$, $l_1-$norm sensitivity is the maximum $l_1-$ norm of $f(D)-f(D^\prime)$, i.e.,
\begin{align}
\Delta f = \max\limits_{D, D^\prime}||f(D)-f(D^\prime)||_1.
\label{equ: sensitivity}
\end{align}
\end{definition}

The Laplace mechanism is commonly used in the literature to achieve differential privacy. 
\begin{definition}{\label{def: Laplace mechanism}\textbf{(Laplace Mechanism) }\cite{dwork2008differential}:}
Given a function $f$: $D \to \mathbb{R}$, the following mechanism $\mathcal{M}$ provides the $\epsilon$-Differential Privacy: 
\begin{align}
\mathcal{M}(D) = f(D) + Lap(\frac{\Delta f}{\epsilon}).
\end{align}
\end{definition}
\subsection{R\'enyi Differential Privacy}

R\'enyi differential privacy (RDP)~\cite{mironov2017renyi} proposed by Ilya Mironov gives a convenient way of tracking cumulative privacy loss when applying differentially private mechanisms. 
RDP measures the R\'enyi divergence~\cite{renyi1961measures} between the output distribution on two neighbouring dataset. 

\begin{definition}{\label{def:Renyi}\textbf{(R\'enyi Divergence)}}~\cite{mironov2017renyi}:
$\mathit{P}$ and $\mathit{Q}$ are probability distribution over $\mathcal{R}$, 
the R\'enyi divergence for $\alpha >1$ is defined as:
\begin{align}
D_{\alpha}(P\parallel Q) \triangleq \frac{1}{\alpha -1}\log{E_{x\sim Q}\left(\frac{P(x)}{Q(x)}\right )^{\alpha}}.
\end{align}
\end{definition}
All the logarithm is natural, $\mathit{P(x)}$ is the density of $\mathit{P}$ at $\mathit{x}$.
\begin{definition}{\label{def:RDP}\textbf{($(\alpha,\epsilon)$-RDP)}}~\cite{mironov2017renyi}:
A randomized mechanism $\mathcal{M}$ gives $(\alpha,\epsilon)$-RDP if for any neighboring datasets $\mathcal{R}$ and $\mathcal{R}^{\prime}$ holds that:
\begin{align}
D_\alpha (\mathcal{M}(\mathcal{R})\parallel \mathcal{M}(\mathcal{R}^{\prime}))\le \epsilon .
\end{align}
\end{definition}
RDP can easily transfer to $(\epsilon,\delta)$-differential private. 
\begin{lemma}{\label{lemma:RDP2GDP}\textbf{(RDP to $(\epsilon, \delta)$-DP)}}~\cite{mironov2017renyi}:
If a mechanism $\mathcal{M}$ satisfies $(\alpha, \epsilon)$-RDP, it also satisfies $(\epsilon+\frac{\log1/\delta}{\alpha-1},\delta)$-DP for any $0<\delta<1$. Besides, $\mathcal{M}$ also satisfies pure $\epsilon$-DP, for $\alpha=\infty$.
\end{lemma}

\subsection{Privacy Amplification by Iteration}

Privacy amplification by iteration has been proven to have its superiority for differentially private optimization algorithms, such as DP-SGD (Differentially-Private Stochastic Gradient Descent)~\cite{Feldman_2018}. The paper reveals the iterative privacy budget under natural settings for DP mechanisms. 

\begin{lemma}{\label{theo:informal}}~\cite{Feldman_2018}:
Assuming $X,X^{\prime} \in \mathcal{R}$, let $\mathcal{M}(X)$ be obtained from $X$ by iterating $T$ times:
\begin{align}
X_{t+1} \doteq f_{t+1}(X_{t})+ m_{t+1}
\end{align}
where $\left \{ f \right \}_{t=1}^{T}$ are contractive maps and $\left \{ m\right \}_{t=1}^{T} \sim \mathcal{N}(0,\sigma ^{2}\mathbb{I}_{d})$. 
Let $\mathcal{M}(X^{\prime})$ be obtained from $X^{\prime}$ under the same process. Then $\mathcal{M}$ satisfies:
\begin{align}
D_\alpha (\mathcal{M}(X)\parallel \mathcal{M}(X^{\prime})) \le \frac{\alpha\left \| X-X^{\prime} \right \|^{2}}{2T\sigma^{2}}.
\end{align}
\end{lemma}
\subsection{Deep Learning in Image Applications}
Deep learning has profoundly changed the landscape of computer vision and image processing. 
The most advanced algorithms in this field are based on DNNs, and most successful DNN architectures use a convolutional structure. Convolutional Neural Network (CNN)~\cite{CNN} consists of neurons with learnable weights and biases. Each neuron receives some input, performs a dot product, and optionally follows its nonlinear activation function. The entire network still implements a distinguishable score function: from the original image pixels on one end to the class score on the other end.

There are several most used CNN architectures. AlexNet~\cite {krizhevsky2012imagenet} was submitted to the ImageNet ILSVRC Challenge in 2012, and its performance far exceeded the second place~\cite {CNN}. It is this work that popularized CNN in computer vision. Szegedy et al. Google proposed GoogLeNet~\cite {szegedy2015going} in 2014. It introduced an Inception module, which significantly reduced the number of parameters in the network (4M, compared to 60 million for AlexNet). GoogLeNet also has multiple subsequent versions, such as Inception-v3~\cite {szegedy2016rethinking}, Inception-v4~\cite {szegedy2017inception}. Karen Simonyan and Andrew Zisserman proposed VGGnet~\cite {Simonyan14c}, which shows that network depth is a key factor for superior performance. ResNet~\cite {he2016deep} was developed by Kaiming He et al. It has a special skip connection and extensive use of batch normalization functions. ResNets are currently widely used in practice.

The image applications include two important categories: 
extracting information from images, 
constructing images with semantic meanings. 
Deep learning has improved the performances from both ends.

For information extraction, 
deep learning has been used for image classification \cite{krizhevsky2012imagenet} \cite{szegedy2017inception} \cite{Simonyan14c},
object detection \cite{ren2015faster}, 
recognition \cite{cimpoi2015deep}, 
tracking \cite{wang2015visual}, 
and semantic segmentation \cite{long2015fully}, etc. 
And it outperforms traditional methods in all these tasks. 
Outputs of DNNs in these applications contain rich information such as type and position of objects, 
identity and action of people, 
thus make DNNs and privacy issues highly relevant.

On the other hand, 
deep learning has also been used to generate synthetic images. 
For example, GAN~\cite{goodfellow2014generative} invented by Ian Goodfellow and his colleagues in 2014 is able to learn to generate new data with the same statistics as the training set. 
Following this initial work, 
some milestone works in GAN were developed such as cGAN~\cite{mirza2014conditional},
StyleGAN~\cite{karras2019style, karras2020analyzing}, etc. 
While GAN could simply generate random new outputs that are similar to the training data, 
Variational Autoencoders (VAEs)~\cite{kingma2013auto} can help to explore variations on data you already have. 
Moreover, 
there have also been researches to combine VAEs and GANs together, 
which can generate compelling results for complex datasets such as images~\cite{mescheder2017adversarial}. 
This VAEs architecture can be used to first take in an input image, 
convert it into a smaller, 
dense representation, 
which the decoder network can use to convert it back to the original image.

\section{Our proposed DP-Image Framework}
\label{sec:System-Model}

\begin{figure*}
\begin{center}
\includegraphics[scale=0.82]{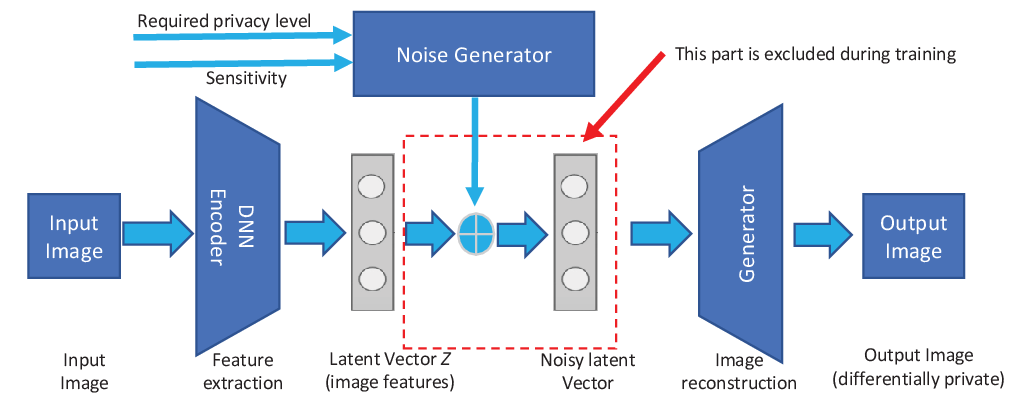}
\end{center}
\caption{Architecture of the DP-Image Framework.}
\label{fig: framework}
\end{figure*}

In this section, 
we first present the adversary model that is considered in this paper, 
and then introduce our proposed DP-Image framework.
\subsection{Adversary Model}
As discussed in Section \ref{sec:Intro}, 
we focus on the mainstream image application: \textit{image publication and sharing scenario}. 
This is a non-interactive setup, 
which includes the case of personal image sharing on social network platforms and commercial image applications such as Google Street View. 

In this scenario, 
the adversary can either manually search or use a web crawler and AI-based tools to automatically identify and collect sensitive information from published images. 
Assisted by the advanced deep learning tools, 
the adversary can obtain different types of information from images, 
for example:
\begin{itemize}
    \item Class or identity of an image using classification or face recognition tools.
    \item Object contained in the image via object detection.
    \item Specific features in the image like numbers or text by natural language processing (NLP) or other feature extractors. 
\end{itemize}

The adversary has two significant differences compared with the traditional database adversary: 
1) he has access to the image; 
2) he extracts information from the image using an advanced query method based on deep learning tools.

On the other hand, 
the users would like to publish or share their photos without personal information leakage. 
Moreover, 
the users would like information coverage satisfies two key criteria: 
1) the images should look realistic and natural after modification, 
so ``Blur'' or ``Mosaic'' is not preferable; 
2) the amount of noise should be controllable, 
so the users can decide the trade-off between privacy and utility.

\subsection{DP-Image Framework: Protecting Image Privacy in Feature Space}
As discussed in Section~\ref{sec:Intro}, 
in the GDPR, 
privacy information is defined as “personal data that are related to an identified or identifiable natural person”. Here, 
the emphasis on privacy is laid on personal identities. 
On the other hand, 
the long history of image processing techniques has proved that the transform domain of an image can provide more useful information than the original pixel domain. 
Also, the adversary can obtain more detailed information by extracting a feature vector $f(X)$ from the image $X$. 
This vector $f(X)$ represents certain features in images. In this case, 
a privacy protection scheme that can blur the feature vector $f(X)$ will be more effective than a scheme that works in the pixel domain.


Based on the above argument, 
we design an image privacy protection framework in feature space. 
The three main modules are feature extraction, noise generator, and image reconstruction, 
as shown in Fig.~\ref{fig: framework}. 
And our motivations for designing these modules in Fig.~\ref{fig: framework} are explained as follows.
\begin{enumerate}
    \item Feature extraction is crucial to feature-level privacy protection.
    The challenge stems from the difficulty in directly extracting the features from pixel values in the original image form, 
    let alone identifying privacy-related features. 
    In order to solve this problem, 
    we propose to use an advanced DNN as the encoder for feature extraction. The output of this encoder will be a feature extraction network that can map an image $X$ to a vector $Z$ in the latent (feature) space.
    \item The noise generator is used to inject noise onto the feature vector. 
    And the amount of noise can be controlled by parameters.
    
    \item The image reconstruction module will transform the perturbed feature vector back to the image domain. 
    This generator can be trained in the same way as in a GAN.
\end{enumerate}

\subsection{DP-Image Definition}
In our \textit{image publication and sharing scenario}, 
the attacker aims to find personal information from the images they have access to. And the image privacy protection aims to ensure that the adversary cannot learn too much about each individual’s private information even though he/she can see the image data. 
This is in stark contrast to the conventional privacy protection methods that might provide an additional layer of protection by not granting the adversary to access the original format of raw data. 

In this sense, 
starting with the DP and RDP notions, 
the definition of image differential privacy can be written as:


\begin{definition}{\label{def:DPImage}\textbf{(($\alpha$,$\epsilon$)-RDP-Image):}} A randomized mechanism $\mathcal{M}$ gives ($\alpha$,$\epsilon$)-RDP-Image if and only if for any two image data $X , X^\prime$, $\mathcal{M}$ satisfies:
\begin{align}
D_\alpha (\mathcal{M}(X)\parallel \mathcal{M}(X^{\prime}))\le \epsilon .
\end{align}
\end{definition}
The above definition is very similar to that of the traditional RDP,  
except that we are enforcing DP for given different input samples, 
instead of requiring there is only a single item (pixel) difference in two images. 
Also, 
we expect the output images of $\mathcal{M}(X)$ and $\mathcal{M}(X^\prime)$ to be indistinguishable from a privacy perspective 
other than in the pixel domain.

Moreover, 
as we plan to add noise in the feature space, 
a generalized sensitivity in RDP of the feature vector $f(X)$ is defined as follows:

\begin{definition}{\label{def:sensitivity}}{\textbf{(Sensitivity in Image Feature Space):}} 
For any two images in the dataset, 
each consisting of $n$ pixels: $X=(x_1,…,x_n)$,  
$X^\prime=(x^\prime_1,…,x^\prime_n)$, 
if $f$ is the function that mapping the image to its feature space, 
the sensitivity $\Delta f$ is defined as the maximum differences in $f(X)$ produced by two different images:
\begin{align}
    \Delta f \doteq \sup_{X,X^{\prime}} ||f(X)-f(X^\prime)||_2.
    \label{eq: sensitivity}
\end{align}
\end{definition}

This indicates the largest difference between the feature vectors of two images.

\subsection{DP-Image Mechanism}

Following Definition~\ref{def:DPImage}, 
to satisfy the DP-Image requirement, 
we implement the DP-Image by add DP noise on the feature space in an iterative manner. 

\begin{algorithm}
\DontPrintSemicolon
\textbf{Parameters: }
Noise coefficient $\sigma$,~
Iteration number $T$.\\

\textbf{Input: } The original image $X$. \\
\textbf{Output: } The released privacy-preserving image $X_{T}$.\\

\For {$0\leq t < T$} {
$z_{t} = f(X_{t})$ \\
$z_{t+1} = f_{c}(z_{t})+\mathcal{N}(0,\sigma ^{2}\mathbb{I}_{d})$ \\
$X_{t+1} = g(z_{t+1})$ \\
}
$X_{T} = X_{t+1}$.

\caption{DP-Image Algorithm by Iteration.}
\label{AlgDPimage}
\end{algorithm}

Where $f(\cdot)$ is a function that maps image $X$ to its feature space vector $z \in \mathbb{R}^m$. 
$f_{c}(\cdot)$ is the clipping function. 
$g(\cdot)$ is a function that maps feature space vector back to image. 
$\mathcal{N}(0,\sigma ^{2}\mathbb{I}_{d})$ are i.i.d random variables drawn from the Gaussian distribution.

It is worth noting that $f_{c}(\cdot)$ is a contractive map, 
which is also said to be 1-Lipschitz. 
\begin{definition}{\label{def:Contractivemaps}\textbf{(Contraction):}} 
For a Banach space ($\mathcal{B}, \left \| \cdot \right \| $), a function $f: \mathcal{B} \to \mathcal{B}$ is said to be contractive (1-Lipschitz) if for all $x,y \in \mathcal{B}$:
\begin{align}
\left \| f(x)-f(y) \right \| \leq \left \| x-y \right \| .
\end{align}
\end{definition}

Based on the iterative DP-Image mechanism above and privacy amplification by iteration (see Lemma~\ref{theo:informal}), we conclude the privacy budget of iterative DP-Image (IDPI).

\begin{theorem}{\label{theo:DPImageIter}(\textbf{Iterative DP-Image (IDPI)})}:
Assuming a start image $X_{0} \in \mathcal{R}$, 
and its corresponding output image $X_{T}$.
$f(\cdot)$ is a map function that maps image into its feature space, 
$f_{c}(\cdot)$ is 1-Lipschitz.
Then, for every $\sigma >0, \alpha>1$,
IDPI($X_{0}, \sigma, T$) satisfies $(\alpha,\frac{\alpha \Delta f^{2}}{2T\sigma^{2}})$-RDP. 
By definition, it also satisfies $(\epsilon+\frac{\log(1/\delta)}{\alpha-1},\delta)$-DP for any $0<\delta<1$, where $\epsilon=\frac{\alpha \Delta f^{2}}{2T\sigma^{2}}$, and satisfies pure $(\frac{\alpha \Delta f^{2}}{2T\sigma^{2}})$-DP, for $\alpha= \infty$. 
\end{theorem}

\begin{proof} Here, 
we give a simplest version of proof. 
First, according to the equation in Lemma~(\ref{theo:informal}). 
$\mathcal{M}(X)$ satisfies $(\alpha,\frac{\alpha \Delta f^{2}}{2T\sigma^{2}})$-RDP:
\begin{align*}
    D_\alpha (\mathcal{M}(f(X))\parallel \mathcal{M}(f(X^{\prime}))) &\leq \frac{\alpha\left \| f(X)-f(X^{\prime}) \right \|^{2}}{2T\sigma^{2}}\\
    &\leq \frac{\alpha \sup ||f(X)-f(X^{\prime})||_{2}^{2}}{2T\sigma^{2}}\\
    &=\frac{\alpha \Delta f^{2}}{2T\sigma^{2}} ~~\text{(Definition~\ref{def:sensitivity})}
\end{align*}
As claimed.

Then the rest of the proof follows from the post-processing property of DP.
Hence, 
we can conclude that if the feature vector is treated with IDPI,
then the reconstructed image $X_{T}$ generated by mechanism $\mathcal{M}$ satisfies the $(\alpha, \epsilon)$-RDP as defined in Definition~\ref{def:DPImage}.

\end{proof}

\section{Experiments}
\label{sec:experiments}

In this section, 
we conduct experiments to validate the effectiveness of the proposed DP-Image scheme. 
As facial recognition is the most widely used privacy sensitive application at present, 
we use face images that contain personal identity information as the dataset in our experiments. 

\subsection{Experiment Setup}

\subsubsection{Dataset}

In our experiment, 
we used the Flickr-Faces-HQ (FFHQ) dataset~\cite{FFHQ}, 
which is a high-quality image dataset of human faces. 
The dataset consists of 70\textit{K} high-quality PNG images with a resolution of $1024\times1024$, 
and contains considerable variation in terms of age, ethnicity, and image background. 

\subsubsection{Evaluation metrics}
\label{subsub: metrics}
We use two groups of metrics to evaluate the performance: 
1) privacy metrics to measure the privacy protection performance: 
including the \textbf{Face Privacy Protection Success Rate (FPPSR)} 
and \textbf{Identity Similarity Score (ISS)}; 
2) utility metrics that can validate the utility of the perturbed images: 
including \textbf{$l_2$-distance}, 
\textbf{Average $l_p$ Distortion (ALD)}, 
\textbf{Structural Similarity (SSIM)}, 
and \textbf{Frechet Inception Distance (FID)}.
\begin{enumerate}
\item{Face privacy protection success rate (FPPSR): 
the average ratio of successful face de-identification. 
It is obtained via using face recognition systems, e.g., ArcFace~\cite{deng2019arcface} and Microsoft Azure Face Recognition API~\cite{AzureFaceAPI} 
to check whether the perturbed image is recognized as a different person from the original image. 
The mechanisms of these two systems are quite similar. 
ArcFace can be treated as a white-box setting as it is open-source, 
while Microsoft API represents a black-box setting.}

\item{Identity Similarity Score (ISS): 
this is also obtained via using face recognition system. 
However, 
rather than using the binary outcome of ``Yes'' or ``No'', 
the soft value that show the similarity between the perturbed image and the original image is used to measure to what degree has the privacy been preserved.}
    \item {Distortion metrics: two distortion metrics are used to measure the amount of noises added to the original image.
    \begin{itemize}
        \item $l_2$-distance computes the Euclidean distance between original and perturbed examples, i.e., $l_2 = \norm {\mathbf Y - \mathbf{X} }_2 $ \\
            \item $ALD_p$~\cite{ling2019deepsec}: $ALD_p=\frac{\left \| \mathbf Y - \mathbf{X} \right \| _{p} }{\left \| \mathbf{X} \right \| _{p} }$.\\
        We use $ALD_{\infty}$ to measure the maximum change in all dimensions of the perturbed images.
    \end{itemize}}
    
\item{SSIM:
SSIM is a method used to measure the similarity between two digital images. 
Compared with the traditional image quality measurement methods, 
such as peak signal-to-noise ratio (PSNR) and mean squared error (MSE), 
SSIM can better match the human judgment of image quality~\cite{sheikh2006statistical}~\cite{wang2009mean}. 
It can be used to quantify the extent that the perturbation is invisible to human eyes. 
A high score means that two images are structurally similar.}
\item{FID~\cite{heusel2017gans}: 
FID captures the similarity between two images based on the deep features calculated using the Google Inception v3 model. 
A lower score indicates that the two images are more similar.}
\end{enumerate}

\subsubsection{Generation of the latent space vector $Z$}

For an image, 
its generated latent space vector $Z$ needs to satisfy two important requirements: 
1) $Z$ is a good representation of the image features; 
2) the original image can be recovered by feeding $Z$ to the generator network.
\begin{figure}[h]
  \begin{center}
  \includegraphics[width=3.3in]{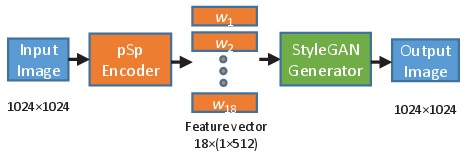}
  \caption{The frame work of generating feature space vector using pSp encoder and reconstructing image with StyleGAN generator.}\label{fig: psp}
  \end{center}
\end{figure}

To achieve this target, 
we use the pSp framework proposed in~\cite{richardson2020encoding}. 
It is an encoder that directly generates a series of style (feature) vectors which can be fed into a pretrained StyleGAN generator. 


As shown in Fig.~\ref{fig: psp}, the dimension of $Z$ generated by the pSp framework is $(512, 18)$, 
consisting of $18$ style vectors, each with a length of $512$ elements. 
These styles are extracted from the corresponding feature map generated from a ResNet backbone, 
where style vectors $(1-3)$ are generated from the small feature map, 
style vectors $(4-7)$ from the medium feature map, 
and style vectors $(8-18)$ from the largest feature map.

\begin{figure}[h]
  \begin{center}
  \includegraphics[width=3 in]{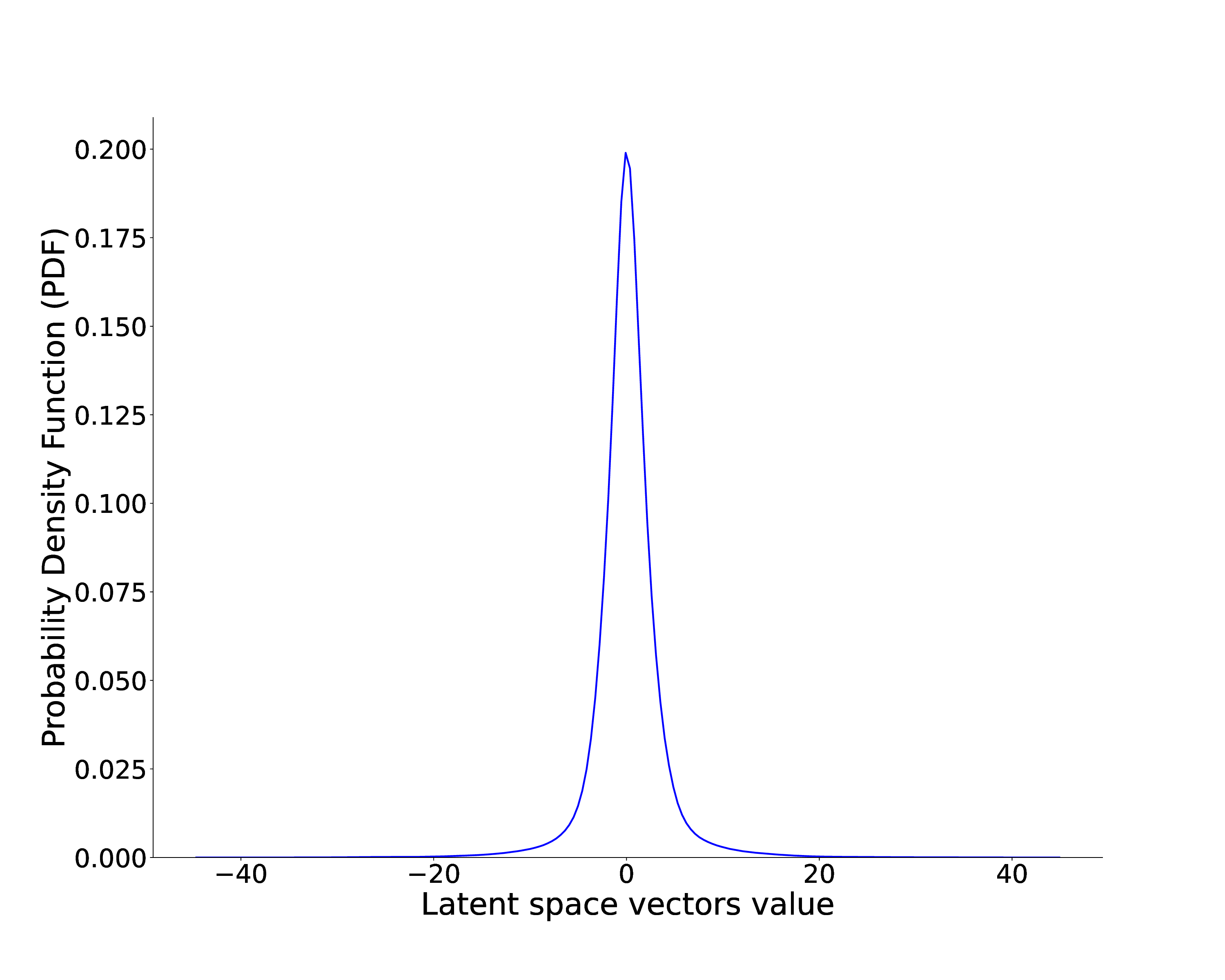}
  \caption{Distribution of the values in latent space feature vector $Z$.}\label{fig: Z_Code_Distribution}
  \end{center}
\end{figure}

An example of the distribution of the generated latent space vector values is exhibited in Fig.~\ref{fig: Z_Code_Distribution}. 
It can be observed that the elements of the $Z$ are in the range of $[-20,20]$, 
with most values near $0$.
In practice, 
the dimension is decided by the feature extraction network. 
Generally speaking, 
the quality of the recovered image improves if we have a higher dimension of $Z$. 
However, 
it comes at the expense of more involved feature extraction of images.

\subsubsection{Image reconstruction using StyleGAN2}

For the image reconstruction, we use the state-of-the-art StyleGAN2~\cite{karras2020analyzing} generator which is trained on the FFHQ dataset. The parameters of the generator are fixed after the training process and the output image is $1024\times1024$. It worth notifying that although StyleGAN2 can accept a simplified  $(512, 1)$ vector as input, it will lose many fine details in high-resolution images~\cite{richardson2020encoding}. Therefore, we use the $(512, 18)$ vector generated by the pSp framework in our experiments.


\subsection{Performance of the DP-Image Mechanism}

\subsubsection{Sensitivity} 
The sensitivity is defined as the maximum element-wise difference of the feature vectors produced by two different images, 
as shown in Eq.~(\ref{eq: sensitivity}). 

In this scenario, 
as the encoder framework is non-convex and does not have a closed-form expression, 
Through observing the distribution of the feature space values (see Fig.~\ref{fig: Z_Code_Distribution}), 
we get an empirical sensitivity by clipping the feature vector into range $[-20,20]$, 
which keeps more than $99\%$ of the image features intact. 
In this case, 
according to Definition~\ref{def:sensitivity}, 
the sensitivity $\Delta f$ can be calculated:
\begin{align}
    \Delta f &\doteq \sup_{X,X^{\prime}} ||f(X)-f(X^\prime)||_2 ~~\text{(Definition~\ref{def:sensitivity})}\\ 
    &= ||20\mathbb{I}_{d},-20\mathbb{I}_{d}||_2\\ 
    &= 3840.
\end{align}

Where $\mathbb{I}_{d}$ is all-ones vector with the same dimension of image feature vector (shape (18,512) in this scheme).

$\Delta f$ scales the noise added to the image features.
By IDPI definition, 
with a large sensitivity, 
the image will need more iterations to achieve a smaller $\epsilon$. 
However, 
from the utility perspective, 
the image could be more real-looking if we adopt a smaller clipping bound.
\subsubsection{Privacy protection performance}

\begin{figure}[htb]
  \begin{center}
  \includegraphics[width=3.2 in]{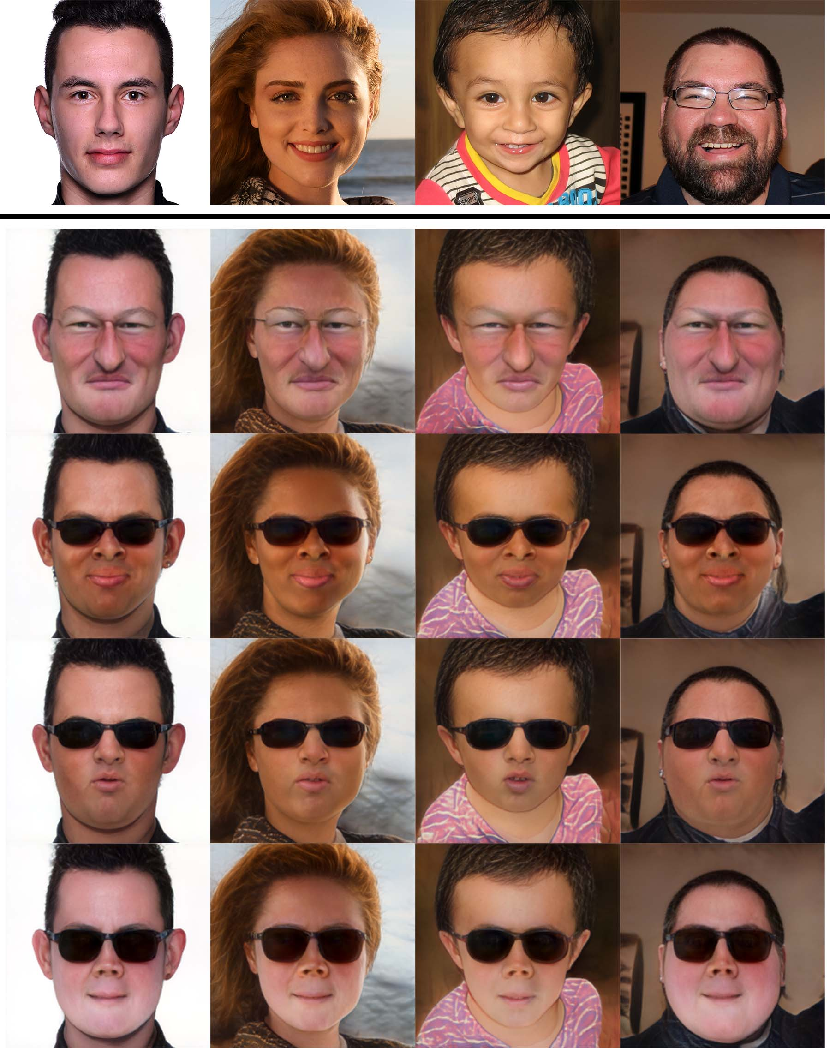}
  \caption{Generated faces with proposed DP-Image method.}
  \label{fig: Example1}
  \end{center}
 \end{figure}
 
\begin{figure*}[htb]
  \begin{center}
  \includegraphics[width=6 in]{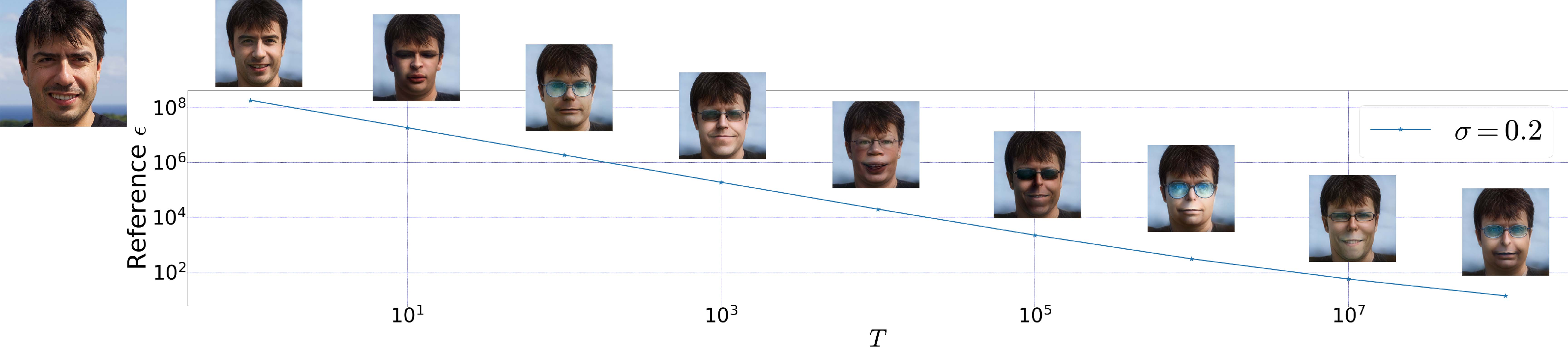}
  \caption{The DP-image visual results with reference $(\epsilon,\delta)$-DP. The x axis and y axis are the image iteration number and reference $\epsilon$ in logarithm. From left to right, they are original image and corresponding generated images with different $T$ and same $\sigma=0.2$. The sensitivity $\Delta f = 3840$. $\delta = 10^{-8}$ by the natural settings on deep learning.}
  \label{fig:R_Epsilon}
  \end{center}
 \end{figure*}

\begin{figure*}[htb]
  \begin{center}
  \includegraphics[width=5 in]{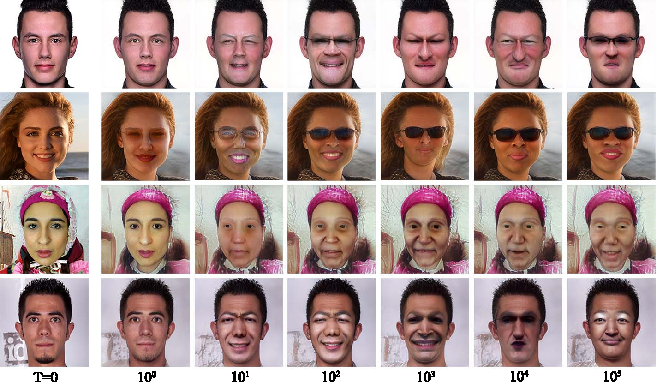}
  \caption{The image visual results for IDPI with different iteration numbers. Above, we give the possible visual results of face images under $\sigma=0.2$ setting. From left to right, the first column are original face images, all other face images are generated with the different iteration number shown above.}
  \label{fig:DiffT}
  \end{center}
 \end{figure*}

To test the performance of our proposed algorithm,
we randomly choose a few images from the dataset and apply our DP-Image scheme on the faces of the image. 
As shown in Fig.~\ref{fig: Example1}, 
the original images are given on the top row, 
all the other four rows are the possible released privacy preserved face images. 
All faces are generate under a setting of IDPI ($X_{0},\sigma = 0.2, T = 10^{4}$). 
In fact, 
for each column, 
there are only one image generated from the original image above, 
i.e. the face identities of the second row are generated from the first face, 
the third row are generated from the second face, and so on. 
From the DP definition, 
all of the four proposed images have the probability to be the 'real' generated image.
In other words, 
the adversary cannot distinguish the queried face by naked eyes, 
even though he is able to access the whole dataset.

Fig.~\ref{fig:R_Epsilon} gives the reference $\epsilon$ under different IDPI iteration numbers.
Next, 
we show a series experiment results to display the visually influence for different IDPI settings. 
Fig.~\ref{fig:DiffT} shows the images with different iteration numbers, 
i.e. IDPI($X_{0},\sigma = 0.2, T = [10^{0},10^{1},10^{2},10^{3},10^{4},10^{5}]$). 
In this case, 
we fixed the $\sigma$ value to observe how the image changes as the iteration number $T$ increases. 
With a larger iteration number, e.g. $10^{5}$, 
the faces does not get much distortion. 
Moreover, 
the IDPI provides good and stable utility during the iteration.
 
\begin{figure}[h]
  \begin{center}
  \includegraphics[width=3.2 in]{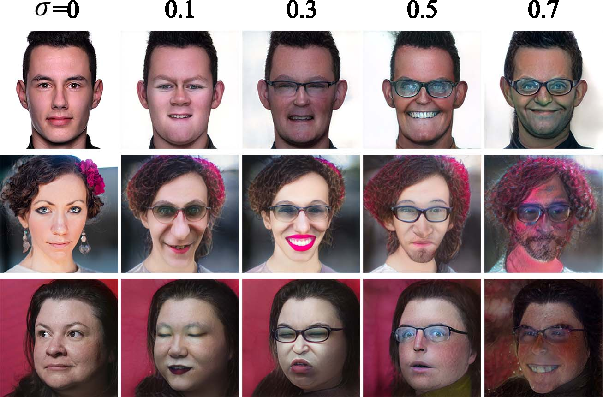}
  \caption{The image visual results for IDPI images with different $\sigma$. Here we give the possible visual results under $T=10^{2}$ setting. From left to right, the first column are original images, all other face images are generated with the different $\sigma$ shown above.}
  \label{fig:Diffsigma}
  \end{center}
 \end{figure}

Then, 
we fix the IDPI parameter $T$ and adjust the privacy budget for each step $\sigma$. 
Fig.~\ref{fig:Diffsigma} shows the images with various $\sigma$ values. 
The faces exhibit poorer utility with a higher $\sigma$. 
Because a larger perturbation in one step will lose too much feature information to maintain the utility of the semantic features. 
Thus, 
$\sigma$ serves as a parameter to control the utility of the whole image.

\begin{figure}[h]
  \begin{center}
  \includegraphics[width=3.25in]{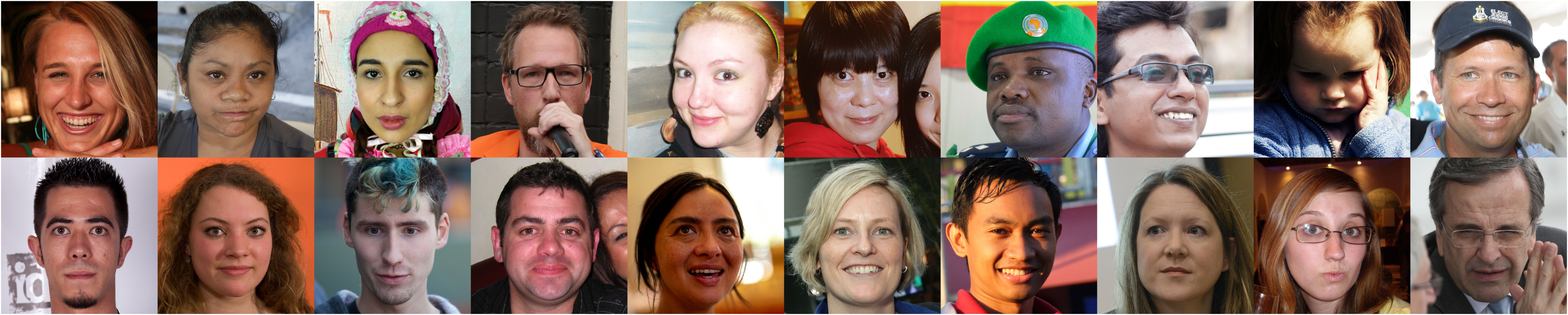}
  \caption{The 20 image examples for the heat map in Fig. \ref{fig: Confidence_Heatmap}. }\label{fig: 20 example}
  \end{center}
\end{figure}

\begin{figure}[h]
  \begin{center}
  \includegraphics[width=2.8 in]{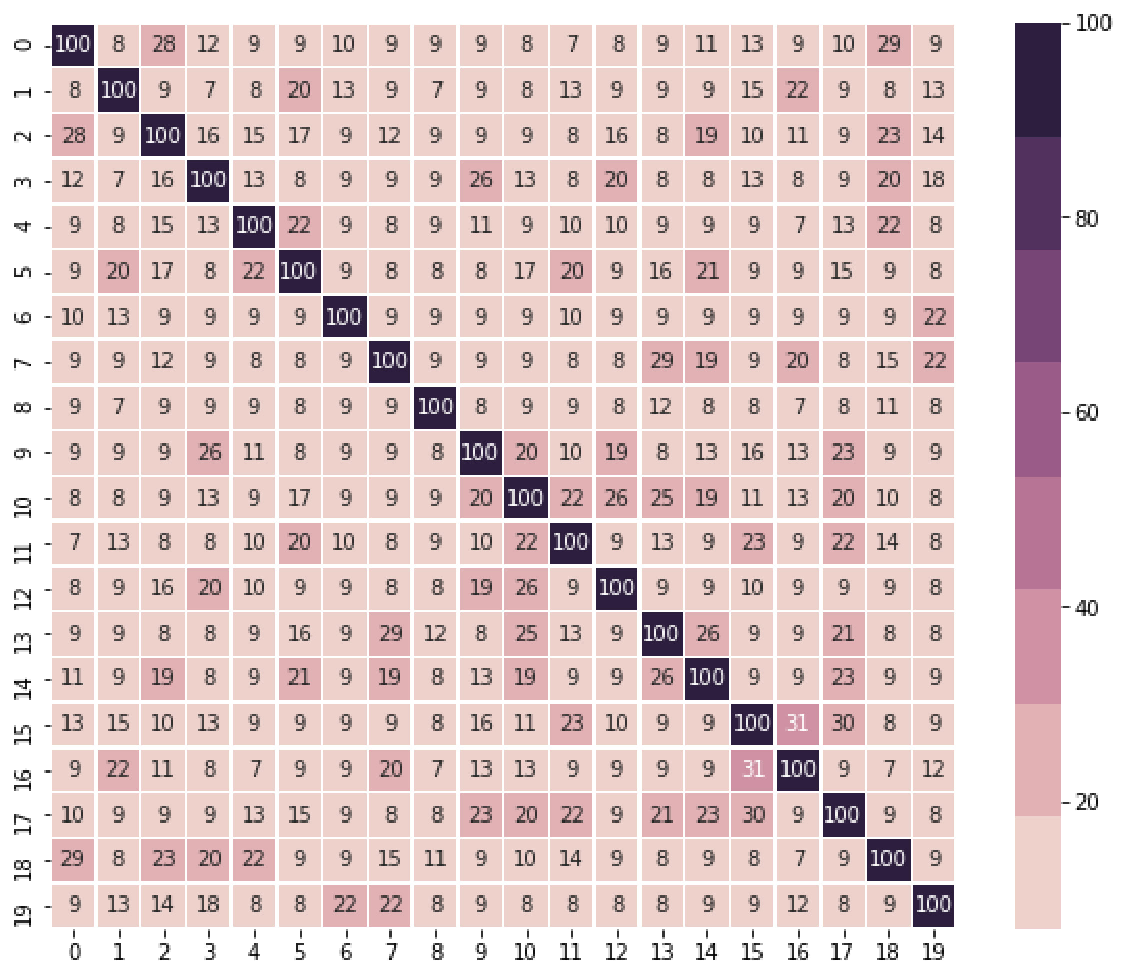}
  \caption{The confidence values of pair-wise image comparison for the 20 example images. Microsoft face API holds high accuracy on face recognition, any two faces with different identity will be give a lower score, e.g. confidence $\leq 50$.}\label{fig: Confidence_Heatmap}
  \end{center}
\end{figure}

To quantifies the performance of our proposed scheme, 
we test the IDPI on the advanced DNN threat model. 
The face recognition models are used to measure the distance of images features. As the DP perturbation are add on the image feature space, the distance measurement on feature space is an essential evaluation in our scheme.
First, 
we use a commercial face recognition system Microsoft Face API as an attacker to evaluate our proposed algorithm.
The Microsoft API can yield a confidence score for any pair of faces being the same person. 
In order to better understand the score of the API, 
we first generate the confidence score values of pair-wise image comparison for the 20 example images 
(see Fig.~\ref{fig: 20 example}) in Fig.~\ref{fig: Confidence_Heatmap}.

\begin{figure*}[h]
     \centering
     \begin{subfigure}[t]{0.35\textwidth}
         \centering
         \includegraphics[width=\textwidth]{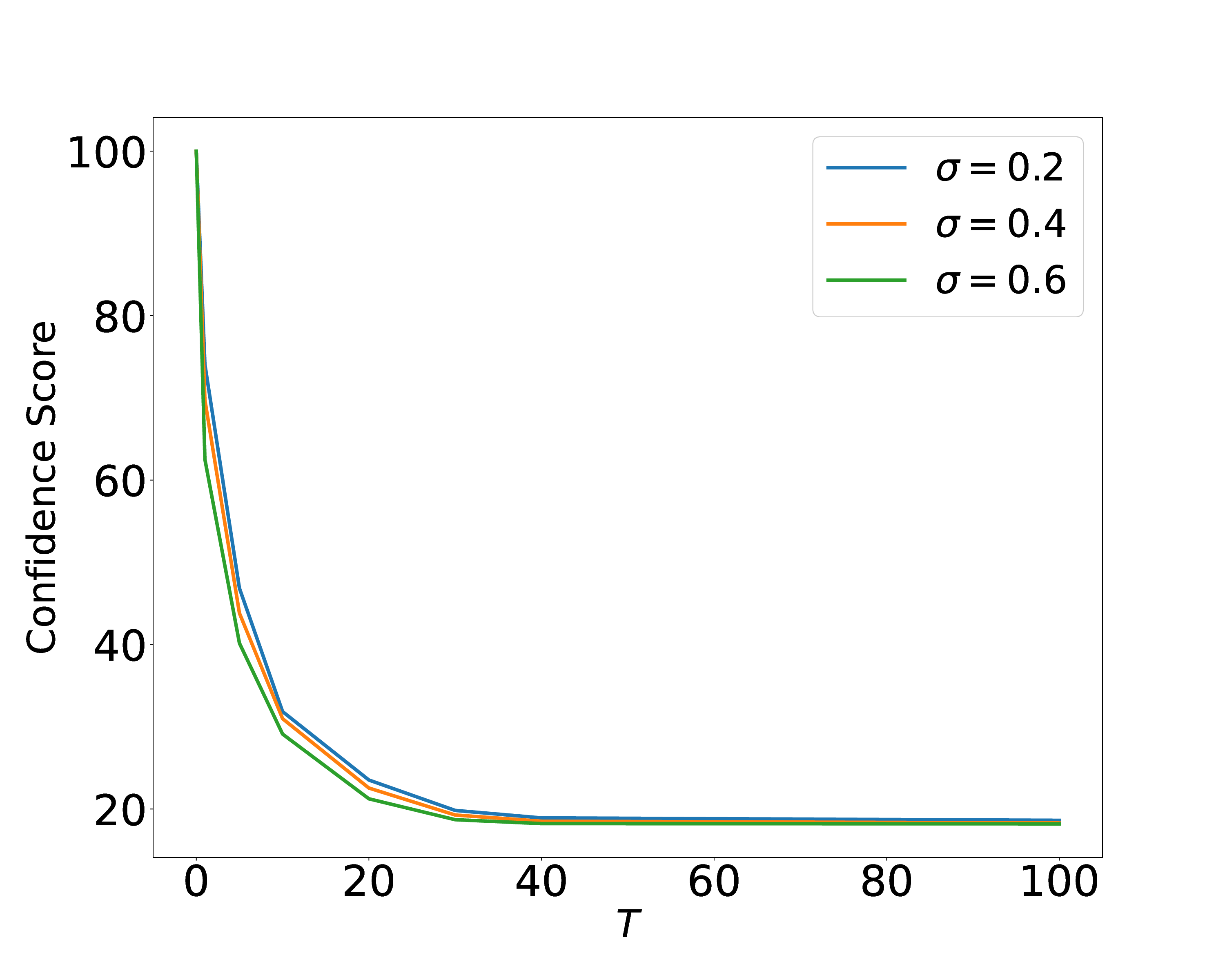}
         \caption{Microsoft Azure Face Recognition API confidence score}
         \label{fig:MSscore}
     \end{subfigure}
     \quad
     \begin{subfigure}[t]{0.35\textwidth}
         \centering
         \includegraphics[width=\textwidth]{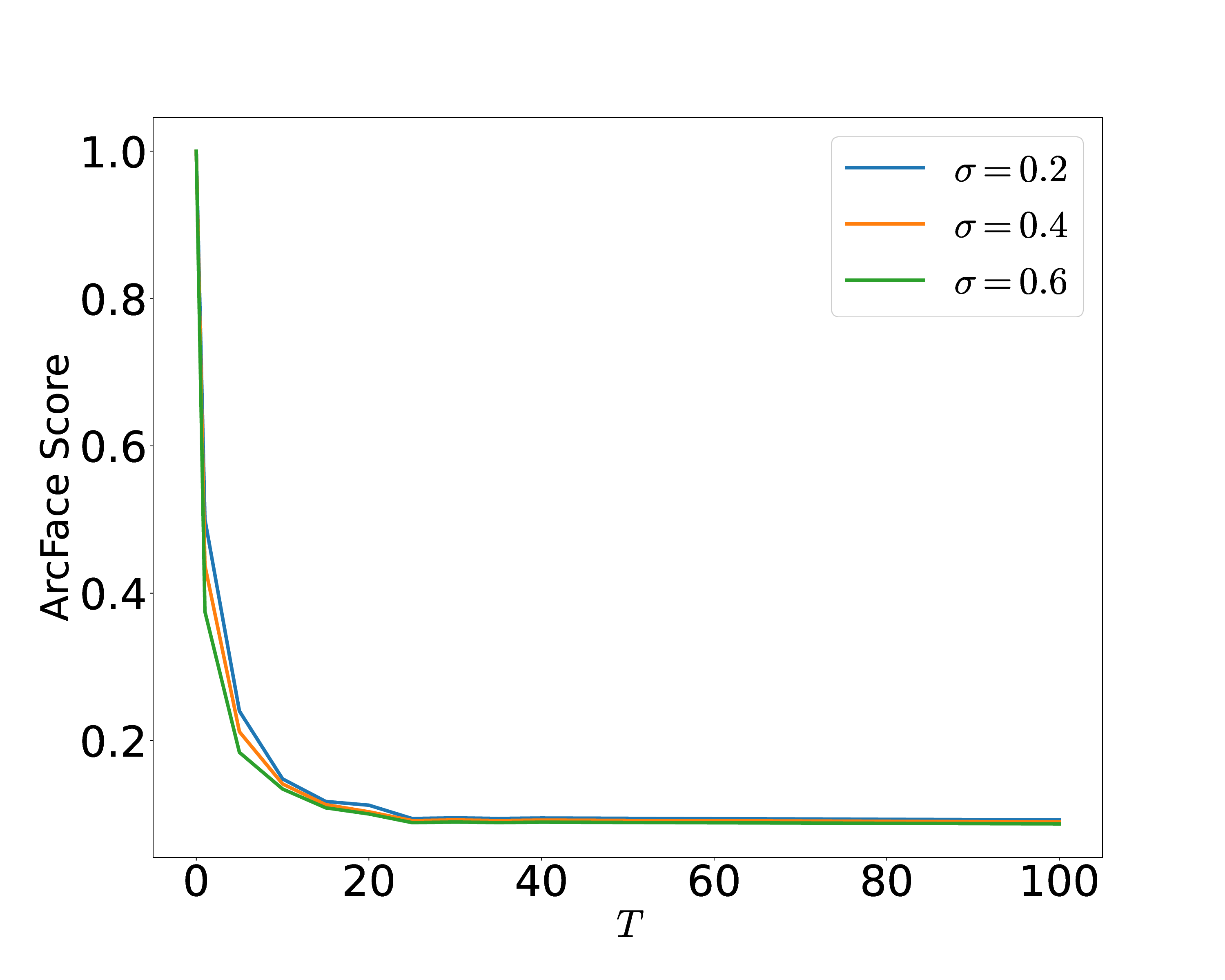}
         \caption{ArcFace score}
         \label{fig:ARCscore}
     \end{subfigure}
        \caption{We evaluate the DP-image performance with Microsoft Face API $(a)$ and the ArcFace $(b)$. $(a)$ The $x$ axis and $y$ axis are iteration number $T$ and Microsoft face API confidence score respectively. $(b)$ The $x$ axis and $y$ axis are iteration number $T$ and ArcFace score respectively. We set $\sigma=0.2, \sigma=0.4, \sigma=0.6$ to compare the confidence score with different setting.}
        \label{fig:Confidence}
\end{figure*} 

\begin{figure*}[h]
    \centering
    \begin{subfigure}[t]{0.35\textwidth}
         \centering
         \includegraphics[width=\textwidth]{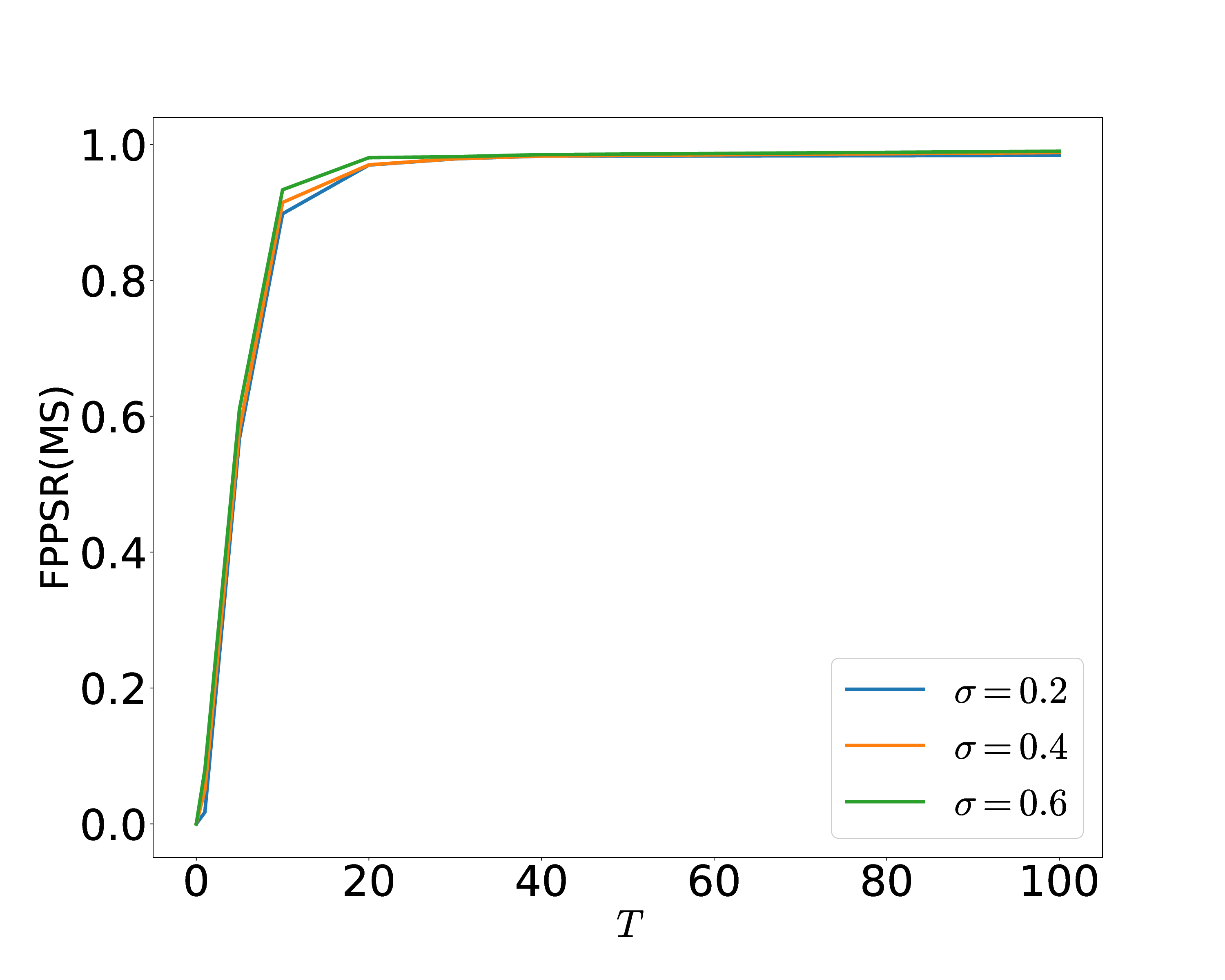}
         \caption{FPPSR of Microsoft Azure Face Recognition API}
         \label{fig:MSFPPSR}
     \end{subfigure}
     \quad
     \begin{subfigure}[t]{0.35\textwidth}
         \centering
         \includegraphics[width=\textwidth]{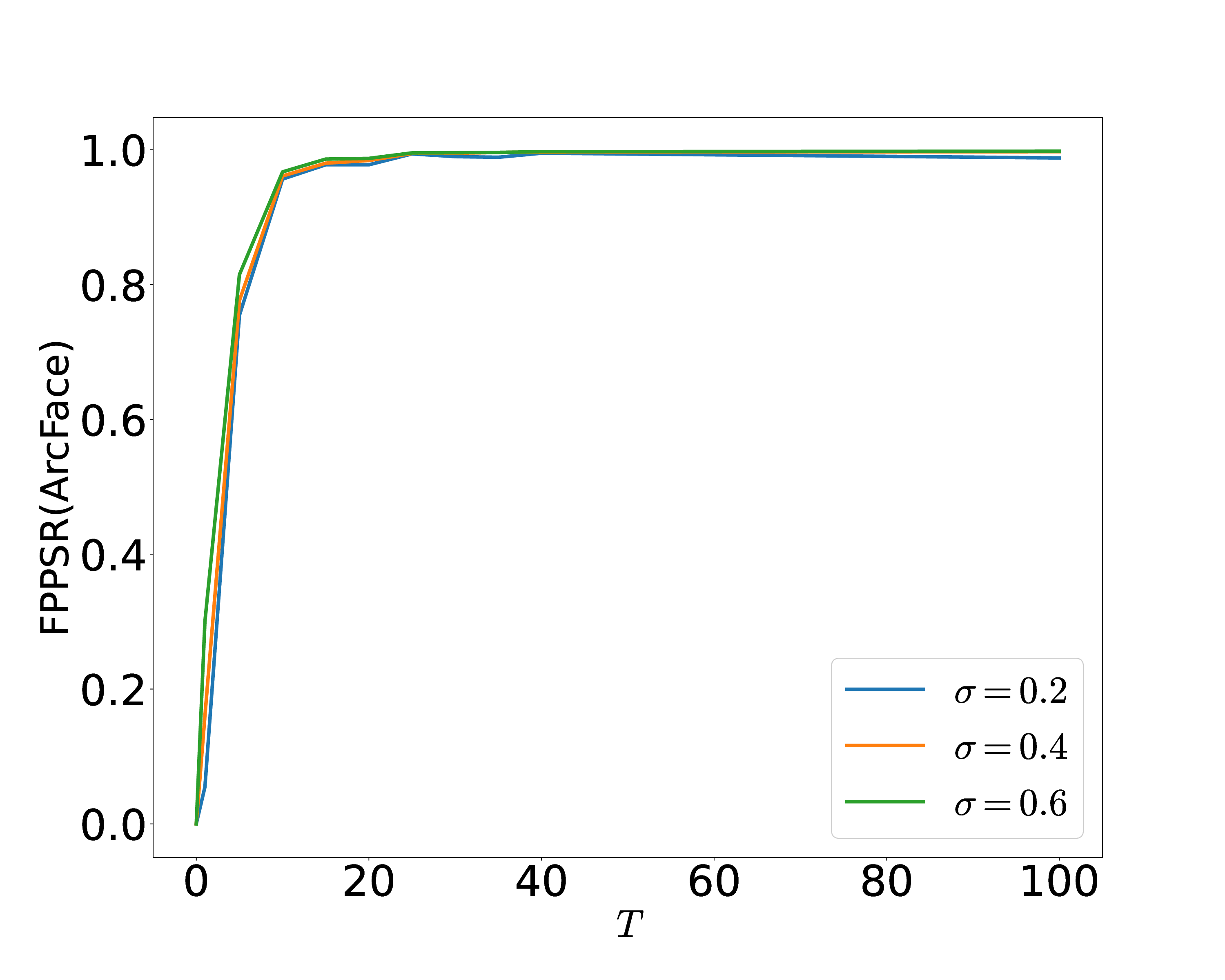}
         \caption{FPPSR of ArcFace}
         \label{fig:ARCFPPSR}
     \end{subfigure}
        \caption{We evaluate the DP-image performance with Microsoft Face API $(a)$ and the ArcFace $(b)$. $(a)$ The $x$ axis and $y$ axis are iteration number $T$ and FPPSR on Microsoft face API respectively. $(b)$ The $x$ axis and $y$ axis are iteration number $T$ and FPPSR on ArcFace respectively. We set $\sigma=0.2, \sigma=0.4, \sigma=0.6$ to compare the FPPSR with different setting.}
        \label{fig:FPPSR}
\end{figure*} 

After that, we make experiments for ISS metric.
In Fig.~\ref{fig:MSscore}, we generate DP-images on FFHQ dataset under different settings and get the Microsoft API confidence scores. The average Microsoft API confidence scores decrease to lower than $50$ within the first $5$ iteration for all $\sigma$. Basically, Microsoft API uses confidence score $50$ as a threshold for image identity. 

Then we attack IDPI with one state-of-the-art face recognition system ArcFace. ArcFace uses DNN (e.g. Resnet) to extract images features and then measure the distance between different features by using Additive Angular Margin Loss. Nowadays, ArcFace is one of the top face recognition model in deep learning area and being implement in many applications. Our results (see Fig.~\ref{fig:ARCscore})show that IDPI can reduce the average ArcFace score to lower than 0.31, which is the empirical threshold in using ArcFace as face identifier, within $5$ iterations for all $\sigma$. 

Besides, the average confidence scores and the ArcFace score are stable at around $18$ and $0.09$ respectively in the following iterations, which indicates a way of choosing better IDPI parameters in practical.

Next, we conducted quantitative evaluation on the dataset using FPPSR privacy evaluation metric introduced in \ref{subsub: metrics}. 
We evaluate our method on both ArcFace and Microsoft Azure Face Recognition APIs. 
We make the same settings with ISS metric and count the success rate by using the threshold introduced in previous experiments (i.e. Microsoft Azure Face Recognition API: $50$, ArcFace: $0.31$). Fig.~\ref{fig:FPPSR} shows that the FPPSR stable at over $99\%$ for both models.  

\subsubsection{Utility}
Table \ref{tab: Compare} compares the utilities of the proposed DP-Image with several traditional perturbation methods when they have approximately the same ISS.
As ArcFace and Microsoft API perform a bit differently, we use both privacy metrics on these methods. 
We try to keep the ISS of ArcFace (i.e. $0.31$) and Microsoft API (i.e. $50$) at a similar level. 
It shows that the proposed DP-Image mechanism can well preserve the perceptual visualization of the image (lowest FID). 
On the other hand, 
the DP-Image method yields a similar performance in the sense of $l_{2}, ALD_{\infty}$ and SSIM,
as these metrics are evaluated on the pixel domain.

\begin{table*}[htb]
\centering
\caption{Utility comparisons of the proposed privacy protection methods with traditional privacy protection methods.}
\label{tab: Compare}
\begin{tabular}{ccccccccc} 
\toprule
     & \multicolumn{4}{c}{ISS(ArcFace 0.31)} & \multicolumn{4}{c}{ISS(Microsoft API 50)}  \\ 
\cline{2-9}
     & $l_2$ & ALD$_{\infty}$ & SSIM & FID                      & $l_2$ & ALD$_{\infty}$ & SSIM & FID   \\ 
\hline
Blur & 71783 & 1.1502  & 0.5895 & 265.8                          & 71533 & 1.1462 & 0.6029 &  220.9                              \\
Mosaic & 70439 & 1.1386 & 0.5705 & 377.5                         & 70439 & 1.1386 & 0.5705 & 377.5                              \\
Adversarial (pixel) & 71657 & 1.1011  & 0.2014 & 391.1                         & 71778 & 1.1029 & 0.2141 & 393.7                              \\
Ours     & 73383 & 1.2567 & 0.5050 &  {\color[HTML]{6434FC} \textbf{155.7}}
                       & 73408 & 1.2531 & 0.5400 &  {\color[HTML]{6434FC} \textbf{152.6}}                              \\
\bottomrule
\end{tabular}
\end{table*}

\section{Discussions}
\label{sec:discussion}
\subsection{Privacy Analysis}
Our experiments in Section~\ref{sec:experiments} show that it is feasible to achieve DP for images in feature space. 
And we can control the amount of noise and the appearance of reconstructed images by adjusting the IDPI parameters. 
Moreover, 
the noise added to the features can not be reverted back so as to reveal the original image, 
thus providing the capability to protect image privacy in a data sharing scenario.

It is worth noting that although IDPI can theoretically achieve strong $\epsilon$-DP (e.g. $\epsilon \leq 10$), the infinite iteration will cost vast of computing power, which naturally decrease the effectiveness of the image. In our experiment, only a relatively small amount of noise can cause the image identity to change. 
There are two major reasons for this phenomena: 
1) although using i.i.d random variable vector $\mathcal{N} = (N_1,...,N_m) $ drawn from the Gaussian distribution can guarantee DP~\cite{Dwork2006calibrating}, 
the bound is quite loose, 
especially when the dimension $m$ is high; 
2) while the DP noise is added on the latent vector $f(X)$, 
we do not need to make $\mathcal{M}(X)$ completely indistinguishable from $\mathcal{M}(X^{\prime})$ in practice. 
Our target is to make the identities of reconstructed images from $X$ and $X^\prime$ close to each other, 
and hence a small variation on the feature vector is enough.

The ultimate purpose of image privacy protection is to prevent de-identification, 
i.e., removing personal identifiers in the image. 
To achieve this target, 
we need to consider two different types of adversary: 
human and AI. 
For a human adversary, 
the generated image needs to satisfy two conditions: 
1) the appearance of the image has changed; 
2) the new image needs to be realistic and natural, 
so that an adversary who has not seen the original image does not know that the new image is a generated instance. 
In this sense, 
our proposed DP-Image method that manipulates images in the feature space is an appealing option. 
For the AI adversary, 
it does not look at the appearance, 
but the features instead. 
In this case, 
the ISS and FPPSR are the primary performance metrics. 
And our experiments show that the DP-Image scheme that adds random noises can generate real-looking face images.

\subsection{Disentangle Identity-related Features in Image}
One interesting discussion is that can we add noises only to the identity-related features in feature space? 
To answer this question, 
the precondition is that we are able to extract the identity-related features from images. 

In terms of the pSp framework that we use in this paper as the encoder, 
we found that the style vectors $(1-7)$ are more identity-related (representing the structure of the faces), 
while the rest of the style vectors $(8-18)$ are not highly related to identity (representing color, pose, etc.)
Therefore, 
we investigate the case of adding noise only to identity-related feature vectors of $Z$, 
which are layers $(1-7)$. 
 
\begin{figure*}[h]
  \begin{center}
  \includegraphics[width=6.5 in]{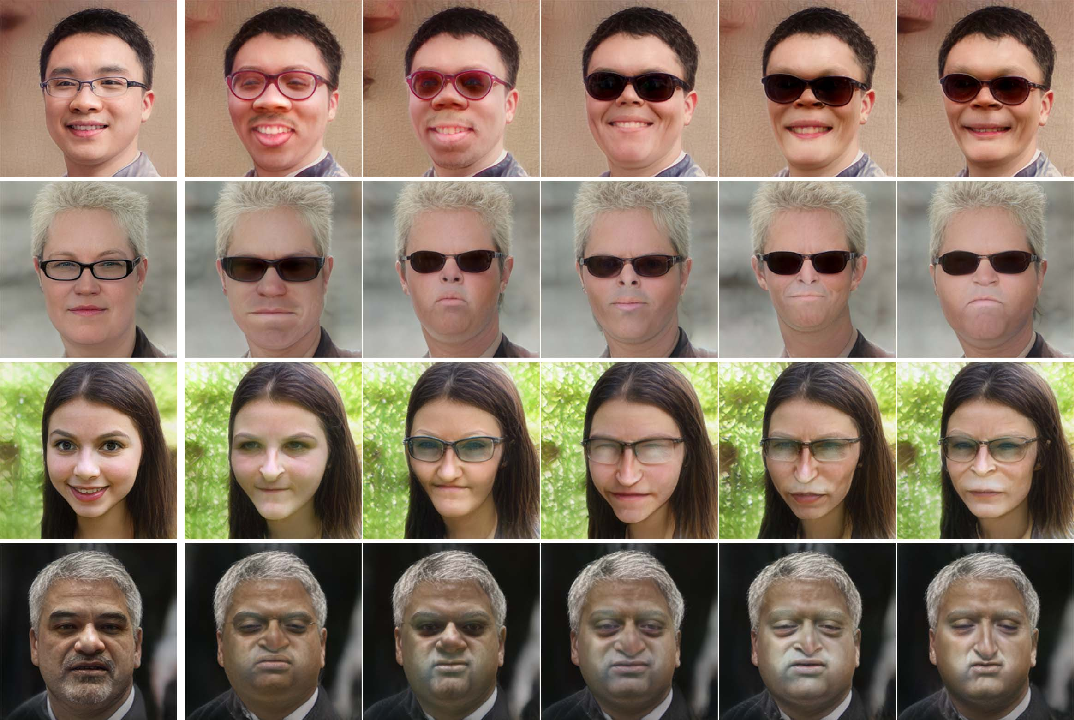}
  \caption{Generated faces with noises added to identity-related features of $Z$. We give the possible visual results of adding noise on image identity features. From left to right, they are original images, the faces protected by IDPI($X_{0}, \sigma = 0.2, T = 10$), IDPI($X_{0}, \sigma = 0.2, T = 20$), IDPI($X_{0}, \sigma = 0.2, T = 30$), IDPI($X_{0}, \sigma = 0.2, T = 40$), IDPI($X_{0}, \sigma = 0.2, T = 50$)}
  \label{fig:z_same_noise}
  \end{center}
 \end{figure*}

As shown in Fig. \ref{fig:z_same_noise}, 
adding noise only to identity-related features in $Z$ could greatly reduces the amount of sensitivity (i.e. from 3840 to 2394 by definition). In the meantime, the effectiveness of privacy persevering keeps the same level.

The pSp framework is not specifically designed for extracting identity-related features, 
therefore the discussions in this subsection is preliminary. 
Extraction of identity-related features is a topic on its own that attracts many recent research interests.
If we can combine the latest extraction technique with the state-of-the-art image autoencoders in the future, 
it will further enhance the performance of the proposed DP-Image mechanism.

\subsection{Image Privacy vs. Database Privacy}

From the experiments, 
we can see that image privacy is quite different from conventional data privacy.
\begin{enumerate}
  \item The protection target of database privacy is the existence of a certain record in the database or not; while image privacy protection targets the identity-related information in the image. 
  \item It is easy to add random noise to the data record and generate meaningful DP perturbed record. 
  However, for an image, 
  adding random noise might lead to distorted and unrealistic pictures. 
  \item The performance of DP-Image needs to be judged by both visual and numerical metrics, 
  as we may need to consider both human and AI as adversaries. 
\end{enumerate}






\section{Related Work}
\label{sec: related work}
As much of the related work on differential privacy and deep learning has been already discussed in Section \ref{sec:Preliminaries}, here we only focus on the works specifically on image privacy protection.

The image privacy issue first attracted people's attention, along with the booming of social networks developing. The proliferation of social networks generated massive photos flooding on the internet that contains sensitive information. For example, Pesce et al. \cite{pesce2012privacy} use photo tags to attack users and get their privacy. The image privacy issue becomes more server with the widely spread of facial recognition systems, as people start to worry that organizations might use their faces for profiling or social control.

The previous mainstream method to combat the image privacy attack is using access control on sensitive contents. For example, Vyas et al. \cite{vyas2009towards} use annotation data to predict the privacy preferences of users and control the shared content. Moreover, Squicciarini et al. \cite{squicciarini2011cope} proposed collaborative privacy management that can let users collaborative control their photos. Similarly, to deal with the privacy issue in facial recognition systems, the current countermeasure is simply banning \cite{Sanfrancisco}. The access control-based method has several limitations. It only gives ``Yes'' or ``No'' options for the use of images, while we need to use part of the information in applications such as Google Street View. And it cannot automate privacy protection based on the privacy information of the image itself, requiring human participation. 

Some more recent image privacy researches focus on changing the identity-related information in images~\cite{wen2020hybrid}. The main techniques are obfuscation and inpainting. Simple obfuscation has been proved to be ineffective against DNN-based recognizer \cite{mcpherson2016defeating, oh2016faceless}. In order to improve the robustness against deep learning aided attack, adversarial example-based methods have been proposed to mislead the neural networks~\cite{oh2017adversarial,liu2017protecting,liu19bmsb, shan2020fawkes, li2019anonymousnet, rajabi2021practicality}. Also, there are some researchers who start to use GAN to generate content to replace the sensitive information in the images~\cite{sun2018natural, wu2019privacy, sun2018hybrid, li2019faceshifter, wang2020infoscrub}. For example, Sun et al. \cite{sun2018natural} proposed GAN-based head inpainting to remove the original identity. Finally, there recently have been a few attempts to combine the DP notion with image privacy. Fan \cite{fan2018image} proposed an $\epsilon$- differential private method in the pixel level of the image. However, making image pixels indistinguishable does not make much sense in practice, and the quality of the generated image is quite low. In another work from the same author \cite{fan2019practical}, metric privacy is defined in the image transformation domain, but the quality of the obfuscated image is still low. Lecuyer et al.~\cite{lecuyer2019certified} proposed the PixelDP framework that includes a DP noise layer in the DNN. PixelDP scheme enforces that the output prediction function is DP provided the input changes on a small number of pixels (when the input is an image). However, the purpose of PixelDP is to increase robustness to adversarial examples, other than image privacy. 


\section{Conclusion and Future Work}
\label{sec: conclusion}

In this paper, 
we have proposed a DP-Image framework IDPI that can protect sensitive personal information in images. 
The major contributions are two-fold.
First, 
we present the novel notion of DP-Image.
Second, 
we propose a method to perturb the image by adding noise to its feature vector in the latent space. 
The effectiveness of the proposed framework is validated by extensive experiments on the face image dataset FFHQ.  

Image privacy and the broader topic of unstructured data privacy is an interesting research direction with many open problems. 
In the future, 
we aim at extending our method in the following directions:
\begin{itemize}
    \item A more comprehensive image feature extraction network that can be applied to different types of images, and even videos.
    \item Investigate how the image feature vector decides the various semantic attributes of images, 
    and whether we can cleanly extract these attributes.
    \item Based on the above understanding, 
    we may be able to concisely control some of the attributes which we believe are crucial from the privacy perspective. 
    \item Investigate DP-Image mechanisms in more complicated images that contain multiple sensitive objects.
\end{itemize}

\bibliographystyle{unsrt}  
\bibliography{DP-Image-ref}

\begin{thebibliography}{10}

\bibitem{facebookprivacy}
Financial Times.
\newblock Facebook privacy breach, 2018.

\bibitem{liu2021ACM}
Bo~Liu, Ming Ding, Sina Shaham, Wenny Rahayu, Farhad Farokhi, and Zihuai Lin.
\newblock When machine learning meets privacy: A survey and outlook.
\newblock {\em ACM Comput. Surv.}, 54(2), March 2021.

\bibitem{Faceapp}
FaceApp.com.
\newblock Faceapp.

\bibitem{mcpherson2016defeating}
Richard McPherson, Reza Shokri, and Vitaly Shmatikov.
\newblock Defeating image obfuscation with deep learning.
\newblock {\em arXiv preprint arXiv:1609.00408}, 2016.

\bibitem{GDPR}
EU.
\newblock The eu general data protection regulation, 2019.

\bibitem{goodfellow2014explaining}
Ian~J Goodfellow, Jonathon Shlens, and Christian Szegedy.
\newblock Explaining and harnessing adversarial examples.
\newblock {\em arXiv preprint arXiv:1412.6572}, 2014.

\bibitem{liu2019adversaries}
Bo~Liu, Ming Ding, Tianqing Zhu, Yong Xiang, and Wanlei Zhou.
\newblock Adversaries or allies? privacy and deep learning in big data era.
\newblock {\em Concurrency and Computation: Practice and Experience},
  31(19):e5102, 2019.

\bibitem{oh2017adversarial}
Seong~Joon Oh, Mario Fritz, and Bernt Schiele.
\newblock Adversarial image perturbation for privacy protection a game theory
  perspective.
\newblock In {\em 2017 IEEE International Conference on Computer Vision
  (ICCV)}, pages 1491--1500. IEEE, 2017.

\bibitem{liu2017protecting}
Yujia Liu, Weiming Zhang, and Nenghai Yu.
\newblock Protecting privacy in shared photos via adversarial examples based
  stealth.
\newblock {\em Security and Communication Networks}, 2017, 2017.

\bibitem{liu19bmsb}
Bo~Liu, Jian Xiong, Yiyan Wu, Ming Ding, and Cynthia~M. Wu.
\newblock Protecting multimedia privacy from both humansand ai.
\newblock In {\em in Proc. IEEE International Symposium on Broadband Multimedia
  Systems and Broadcasting}, 2019.

\bibitem{shan2020fawkes}
Shawn Shan, Emily Wenger, Jiayun Zhang, Huiying Li, Haitao Zheng, and Ben~Y
  Zhao.
\newblock Fawkes: protecting privacy against unauthorized deep learning models.
\newblock In {\em 29th USENIX Security Symposium (USENIX Security 20)}, pages
  1589--1604, 2020.

\bibitem{li2019anonymousnet}
Tao Li and Lei Lin.
\newblock Anonymousnet: Natural face de-identification with measurable privacy.
\newblock In {\em Proceedings of the IEEE Conference on Computer Vision and
  Pattern Recognition Workshops}, pages 0--0, 2019.

\bibitem{rajabi2021practicality}
Arezoo Rajabi, Rakesh~B Bobba, Mike Rosulek, Charles~V Wright, and Wu-chi Feng.
\newblock On the (im) practicality of adversarial perturbation for image
  privacy.
\newblock {\em Proceedings on Privacy Enhancing Technologies}, 2021(1):85--106,
  2021.

\bibitem{sun2018natural}
Qianru Sun, Liqian Ma, Seong Joon~Oh, Luc Van~Gool, Bernt Schiele, and Mario
  Fritz.
\newblock Natural and effective obfuscation by head inpainting.
\newblock In {\em Proceedings of the IEEE Conference on Computer Vision and
  Pattern Recognition}, pages 5050--5059, 2018.

\bibitem{wu2019privacy}
Yifan Wu, Fan Yang, Yong Xu, and Haibin Ling.
\newblock Privacy-protective-gan for privacy preserving face de-identification.
\newblock {\em Journal of Computer Science and Technology}, 34(1):47--60, 2019.

\bibitem{sun2018hybrid}
Qianru Sun, Ayush Tewari, Weipeng Xu, Mario Fritz, Christian Theobalt, and
  Bernt Schiele.
\newblock A hybrid model for identity obfuscation by face replacement.
\newblock In {\em Proceedings of the European Conference on Computer Vision
  (ECCV)}, pages 553--569, 2018.

\bibitem{li2019faceshifter}
Lingzhi Li, Jianmin Bao, Hao Yang, Dong Chen, and Fang Wen.
\newblock Faceshifter: Towards high fidelity and occlusion aware face swapping.
\newblock {\em arXiv preprint arXiv:1912.13457}, 2019.

\bibitem{wang2020infoscrub}
Hui-Po Wang, Tribhuvanesh Orekondy, and Mario Fritz.
\newblock Infoscrub: Towards attribute privacy by targeted obfuscation.
\newblock {\em arXiv preprint arXiv:2005.10329}, 2020.

\bibitem{fan2018image}
Liyue Fan.
\newblock Image pixelization with differential privacy.
\newblock In {\em IFIP Annual Conference on Data and Applications Security and
  Privacy}, pages 148--162. Springer, 2018.

\bibitem{fan2019practical}
Liyue Fan.
\newblock Practical image obfuscation with provable privacy.
\newblock In {\em 2019 IEEE International Conference on Multimedia and Expo
  (ICME)}, pages 784--789. IEEE, 2019.

\bibitem{lecuyer2019certified}
Mathias Lecuyer, Vaggelis Atlidakis, Roxana Geambasu, Daniel Hsu, and Suman
  Jana.
\newblock Certified robustness to adversarial examples with differential
  privacy.
\newblock In {\em 2019 IEEE Symposium on Security and Privacy (SP)}, pages
  656--672. IEEE, 2019.

\bibitem{dwork2008differential}
Cynthia Dwork.
\newblock Differential privacy: A survey of results.
\newblock In {\em International Conference on Theory and Applications of Models
  of Computation}, pages 1--19. Springer, 2008.

\bibitem{mironov2017renyi}
Ilya Mironov.
\newblock R{\'e}nyi differential privacy.
\newblock In {\em 2017 IEEE 30th Computer Security Foundations Symposium
  (CSF)}, pages 263--275. IEEE, 2017.

\bibitem{Dwork2006calibrating}
Cynthia Dwork, Frank McSherry, Kobbi Nissim, and Adam Smith.
\newblock Calibrating noise to sensitivity in private data analysis.
\newblock In {\em Proceedings of the theory of cryptography conference
  (TCC'06)}, pages 265--284. Springer, 2006.

\bibitem{renyi1961measures}
Alfr{\'e}d R{\'e}nyi et~al.
\newblock On measures of entropy and information.
\newblock In {\em Proceedings of the Fourth Berkeley Symposium on Mathematical
  Statistics and Probability, Volume 1: Contributions to the Theory of
  Statistics}. The Regents of the University of California, 1961.

\bibitem{Feldman_2018}
Vitaly Feldman, Ilya Mironov, Kunal Talwar, and Abhradeep Thakurta.
\newblock Privacy amplification by iteration.
\newblock {\em 2018 IEEE 59th Annual Symposium on Foundations of Computer
  Science (FOCS)}, Oct 2018.

\bibitem{CNN}
Sung-Soo Kim.
\newblock Convolutional neural networks for visual recognition, 2016.

\bibitem{krizhevsky2012imagenet}
Alex Krizhevsky, Ilya Sutskever, and Geoffrey~E Hinton.
\newblock Imagenet classification with deep convolutional neural networks.
\newblock In {\em Advances in neural information processing systems}, pages
  1097--1105, 2012.

\bibitem{szegedy2015going}
Christian Szegedy, Wei Liu, Yangqing Jia, Pierre Sermanet, Scott Reed, Dragomir
  Anguelov, Dumitru Erhan, Vincent Vanhoucke, Andrew Rabinovich, et~al.
\newblock Going deeper with convolutions.
\newblock Cvpr, 2015.

\bibitem{szegedy2016rethinking}
Christian Szegedy, Vincent Vanhoucke, Sergey Ioffe, Jon Shlens, and Zbigniew
  Wojna.
\newblock Rethinking the inception architecture for computer vision.
\newblock In {\em Proceedings of the IEEE Conference on Computer Vision and
  Pattern Recognition}, pages 2818--2826, 2016.

\bibitem{szegedy2017inception}
Christian Szegedy, Sergey Ioffe, Vincent Vanhoucke, and Alexander~A Alemi.
\newblock Inception-v4, inception-resnet and the impact of residual connections
  on learning.
\newblock In {\em AAAI}, volume~4, page~12, 2017.

\bibitem{Simonyan14c}
K.~Simonyan and A.~Zisserman.
\newblock Very deep convolutional networks for large-scale image recognition.
\newblock {\em CoRR}, abs/1409.1556, 2014.

\bibitem{he2016deep}
Kaiming He, Xiangyu Zhang, Shaoqing Ren, and Jian Sun.
\newblock Deep residual learning for image recognition.
\newblock In {\em Proceedings of the IEEE Conference on Computer Vision and
  Pattern Recognition}, pages 770--778, 2016.

\bibitem{ren2015faster}
Shaoqing Ren, Kaiming He, Ross Girshick, and Jian Sun.
\newblock Faster r-cnn: Towards real-time object detection with region proposal
  networks.
\newblock In {\em Advances in neural information processing systems}, pages
  91--99, 2015.

\bibitem{cimpoi2015deep}
Mircea Cimpoi, Subhransu Maji, and Andrea Vedaldi.
\newblock Deep filter banks for texture recognition and segmentation.
\newblock In {\em Proceedings of the IEEE Conference on Computer Vision and
  Pattern Recognition}, pages 3828--3836, 2015.

\bibitem{wang2015visual}
Lijun Wang, Wanli Ouyang, Xiaogang Wang, and Huchuan Lu.
\newblock Visual tracking with fully convolutional networks.
\newblock In {\em Proceedings of the IEEE international conference on computer
  vision}, pages 3119--3127, 2015.

\bibitem{long2015fully}
Jonathan Long, Evan Shelhamer, and Trevor Darrell.
\newblock Fully convolutional networks for semantic segmentation.
\newblock In {\em Proceedings of the IEEE Conference on Computer Vision and
  Pattern Recognition}, pages 3431--3440, 2015.

\bibitem{goodfellow2014generative}
Ian Goodfellow, Jean Pouget-Abadie, Mehdi Mirza, Bing Xu, David Warde-Farley,
  Sherjil Ozair, Aaron Courville, and Yoshua Bengio.
\newblock Generative adversarial nets.
\newblock In {\em Advances in neural information processing systems}, pages
  2672--2680, 2014.

\bibitem{mirza2014conditional}
Mehdi Mirza and Simon Osindero.
\newblock Conditional generative adversarial nets.
\newblock {\em arXiv preprint arXiv:1411.1784}, 2014.

\bibitem{karras2019style}
Tero Karras, Samuli Laine, and Timo Aila.
\newblock A style-based generator architecture for generative adversarial
  networks.
\newblock In {\em Proceedings of the IEEE Conference on Computer Vision and
  Pattern Recognition}, pages 4401--4410, 2019.

\bibitem{karras2020analyzing}
Tero Karras, Samuli Laine, Miika Aittala, Janne Hellsten, Jaakko Lehtinen, and
  Timo Aila.
\newblock Analyzing and improving the image quality of stylegan.
\newblock In {\em Proceedings of the IEEE/CVF Conference on Computer Vision and
  Pattern Recognition}, pages 8110--8119, 2020.

\bibitem{kingma2013auto}
Diederik~P Kingma and Max Welling.
\newblock Auto-encoding variational bayes.
\newblock {\em arXiv preprint arXiv:1312.6114}, 2013.

\bibitem{mescheder2017adversarial}
Lars Mescheder, Sebastian Nowozin, and Andreas Geiger.
\newblock Adversarial variational bayes: Unifying variational autoencoders and
  generative adversarial networks.
\newblock In {\em Proceedings of the 34th International Conference on Machine
  Learning-Volume 70}, pages 2391--2400. JMLR. org, 2017.

\bibitem{FFHQ}
NVIDIA.
\newblock Flickr-faces-hq (ffhq) dataset, 2019.

\bibitem{deng2019arcface}
Jiankang Deng, Jia Guo, Niannan Xue, and Stefanos Zafeiriou.
\newblock Arcface: Additive angular margin loss for deep face recognition.
\newblock In {\em Proceedings of the IEEE Conference on Computer Vision and
  Pattern Recognition}, pages 4690--4699, 2019.

\bibitem{AzureFaceAPI}
Microsoft.
\newblock Azure facial recognition api.

\bibitem{ling2019deepsec}
Xiang Ling, Shouling Ji, Jiaxu Zou, Jiannan Wang, Chunming Wu, Bo~Li, and Ting
  Wang.
\newblock Deepsec: A uniform platform for security analysis of deep learning
  model.

\bibitem{sheikh2006statistical}
Hamid~R Sheikh, Muhammad~F Sabir, and Alan~C Bovik.
\newblock A statistical evaluation of recent full reference image quality
  assessment algorithms.
\newblock {\em IEEE Transactions on image processing}, 15(11):3440--3451, 2006.

\bibitem{wang2009mean}
Zhou Wang and Alan~C Bovik.
\newblock Mean squared error: Love it or leave it? a new look at signal
  fidelity measures.
\newblock {\em IEEE signal processing magazine}, 26(1):98--117, 2009.

\bibitem{heusel2017gans}
Martin Heusel, Hubert Ramsauer, Thomas Unterthiner, Bernhard Nessler, and Sepp
  Hochreiter.
\newblock Gans trained by a two time-scale update rule converge to a local nash
  equilibrium.
\newblock In {\em Advances in neural information processing systems}, pages
  6626--6637, 2017.

\bibitem{richardson2020encoding}
Elad Richardson, Yuval Alaluf, Or~Patashnik, Yotam Nitzan, Yaniv Azar, Stav
  Shapiro, and Daniel Cohen-Or.
\newblock Encoding in style: a stylegan encoder for image-to-image translation.
\newblock {\em arXiv preprint arXiv:2008.00951}, 2020.

\bibitem{pesce2012privacy}
Jo{\~a}o~Paulo Pesce, Diego~Las Casas, Gustavo Rauber, and Virg{\'\i}lio
  Almeida.
\newblock Privacy attacks in social media using photo tagging networks: a case
  study with facebook.
\newblock In {\em Proceedings of the 1st Workshop on Privacy and Security in
  Online Social Media}, pages 1--8, 2012.

\bibitem{vyas2009towards}
Nitya Vyas, Anna~C Squicciarini, Chih-Cheng Chang, and Danfeng Yao.
\newblock Towards automatic privacy management in web 2.0 with semantic
  analysis on annotations.
\newblock In {\em 2009 5th International Conference on Collaborative Computing:
  Networking, Applications and Worksharing}, pages 1--10. IEEE, 2009.

\bibitem{squicciarini2011cope}
Anna~C Squicciarini, Heng Xu, and Xiaolong Zhang.
\newblock Cope: Enabling collaborative privacy management in online social
  networks.
\newblock {\em Journal of the American Society for Information Science and
  Technology}, 62(3):521--534, 2011.

\bibitem{Sanfrancisco}
The New~York Times.
\newblock {San Francisco Bans Facial Recognition Technology}, 2019.

\bibitem{wen2020hybrid}
Yunqian Wen, Bo~Liu, Rong Xie, Yunhui Zhu, Jingyi Cao, and Li~Song.
\newblock A hybrid model for natural face de-identiation with adjustable
  privacy.
\newblock In {\em 2020 IEEE International Conference on Visual Communications
  and Image Processing (VCIP)}, pages 269--272. IEEE, 2020.

\bibitem{oh2016faceless}
Seong~Joon Oh, Rodrigo Benenson, Mario Fritz, and Bernt Schiele.
\newblock Faceless person recognition: Privacy implications in social media.
\newblock In {\em European Conference on Computer Vision}, pages 19--35.
  Springer, 2016.

\end{thebibliography}






\end{document}